\documentclass[pra,floatfix,notitlepage,twocolumn,showpacs,preprintnumbers,longbibliography,nofootinbi]{revtex4-1}

\usepackage[a4paper,left=1cm,right=1cm,top=2.5cm,bottom=2.5cm]{geometry}

\usepackage[latin1]{inputenc}

\usepackage[colorlinks=true,citecolor=NavyBlue,urlcolor=NavyBlue,linkcolor=NavyBlue]{hyperref}
\usepackage[cm]{fullpage}
\usepackage{graphicx}
\usepackage[english]{babel}
\usepackage{amsmath}
\usepackage{amssymb}
\usepackage{cancel}
\usepackage{natbib}
\usepackage{comment}
\usepackage{color}
\usepackage[normalem]{ulem}
\usepackage{bbold}
\usepackage{physics} 
\usepackage[dvipsnames]{xcolor}
\usepackage{blindtext}
\usepackage{makeidx}
\usepackage{etoolbox}
\usepackage{titletoc} 

\allowdisplaybreaks

\newcommand{\T}[1]{\text{#1}}

\newcommand{\stkout}[1]{\ifmmode\text{\sout{\ensuremath{#1}}}\else\sout{#1}\fi}

\begin{document}
	\raggedbottom

	\title{
		Many-Body Quantum Optics in a Bose-Hubbard Waveguide
	}

	\author{Federico Roccati}
	\affiliation{Department of Physics, Columbia University, New York, New York 10027, USA,}
	\affiliation{Max Planck Institute for the Science of Light, 91058 Erlangen, Germany}
	\affiliation{Quantum Theory Group, Dipartimento di Fisica e Chimica Emilio Segr\`e,
		Universit\`a degli Studi di Palermo, via Archirafi 36, I-90123 Palermo, Italy}
	
		\begin{abstract}

		Waveguide quantum electrodynamics (QED) studies the interaction between quantum emitters and guided photons in one-dimension. When the waveguide hosts interacting photons, it becomes a platform to explore many-body quantum optics. 
		However, the influence of photonic correlations on emitter dynamics remains poorly understood.
		In this work, we study the collective decay and coherent interactions of quantum emitters coupled to a one-dimensional Bose-Hubbard waveguide, an array of coupled photonic modes with repulsive on-site interactions that supports superfluid and Mott insulating phases.
		We show that photon-photon interactions alone can trigger a superradiant burst, independent of emitter spacing and transition frequency.
			In the off-resonant regime, emitters exhibit two distinct types of mediated interactions. Delocalized superfluid excitations yield persistent, long-range couplings, challenging the noninteracting paradigm, where long-range (dissipative) couplings occur in-band, whereas  short-range and coherent interactions occur in a bandgap. In the Mott limit, quasiparticles generate short-range interactions mediated by doublons and holons.
		Our work bridges many-body physics and waveguide QED, revealing how photonic many-body states shape emitter dynamics.
	\end{abstract}
	
	\maketitle
	
	{\let\newpage\relax\maketitle}

	\section{Introduction}
	Waveguide quantum electrodynamics (QED)~\cite{JohnPRL1990,JohnPRA1994,ShenPRL2005,ShenPRA2009,GobanPRL2015,SipahigilScience2016,CorzoNature2019,FosterPRL2019,MirhosseiniNature2019,KannanNature2020,SheremetRMP2023,CiccarelloOptPhotonNews2024} has emerged as a powerful paradigm for engineering light-matter interactions, where emitters coupled to confined photonic modes exhibit collective phenomena like superradiance~\cite{DickePR1954,GrossPhysicsReports1982,AsenjoGarciaPRX2017,CardenasPRL2023}, subradiance~\cite{AlbrechtNJP2019}, and photon-mediated spin models~\cite{TudelaPRL2017,Tudela2017b,ShiNJP2018,BelloSciAdv2019,SanchezPRA20,KimPRX2021,ScigliuzzoPRX2022,LeonforteQST2025,Benedetto2025dipoledipole}. In traditional setups, the photonic bath is assumed to be non-interacting, enabling exact solutions. Recent advances, however, have begun to explore nonlinear photonic environments, where photon-photon interactions introduce correlations that modify emitter dynamics. For instance, attractive interactions can lead to bound photon states~\cite{WangPRRes2024} or supercorrelated decay~\cite{WangPRL2020}.
	
	Yet, the role of repulsive photon-photon interactions--ubiquitous in quantum optical platforms~\cite{KapitPRX2014,RoushanScience2017,MaNature2019,CarusottoNatPhys2020,ScigliuzzoPRX2022}--remains largely unexplored in waveguide QED. In the many-body regime, repulsive interactions drive quantum phases, such as the superfluid-to-Mott insulator transition~\cite{FisherPRB1989,SheshadriEPL1993,FreericksEPL1994,JakschPRL1998,ElstnerPRB1999,GreinerNature2002,GreentreeNatPhys2006}, whose quasiparticle excitations (Bogoliubov modes~\cite{vanOostenPRA2001} or doublons/holons~\cite{BarmettlerPRA2012}) could mediate novel emitter-emitter couplings. Prior works in nonlinear waveguide QED focused either on waveguides with attractive photon-photon interactions~\cite{WangPRL2020,TalukdarPRA2022,WangPRRes2024}, nonlinear waveguides with parametric gain~\cite{KarnieliArXiv2024}, or driven-dissipative many-body systems~\cite{CaleffiPRL2023}, leaving open how many-body photonic states influence collective emitter phenomena.
	
	Here, we bridge this gap by studying a systems made of quantum emitters coupled to a Bose-Hubbard waveguide--a one-dimensional (1D) lattice hosting repulsive photon-photon interactions. We uncover two key results: (1) The superradiant burst--a hallmark of collective decay--becomes tunable via the photon-photon interaction strength, even for fixed emitter spacing and frequency. (2) Off-resonant emitters develop dipole-dipole interactions mediated by many-body quasiparticles: delocalized Bogoliubov modes in the superfluid phase induce distance-insensitive couplings (in stark contrast with short-range bandgap photon-mediated interactions), while Mott-insulator excitations recover standard waveguide-QED-like behavior, but with doublon/holon dressing. Our results establish a new avenue for exploring the interplay between waveguide QED and many-body physics.
	
	\begin{figure}
		\centering
		\includegraphics[width=\columnwidth]{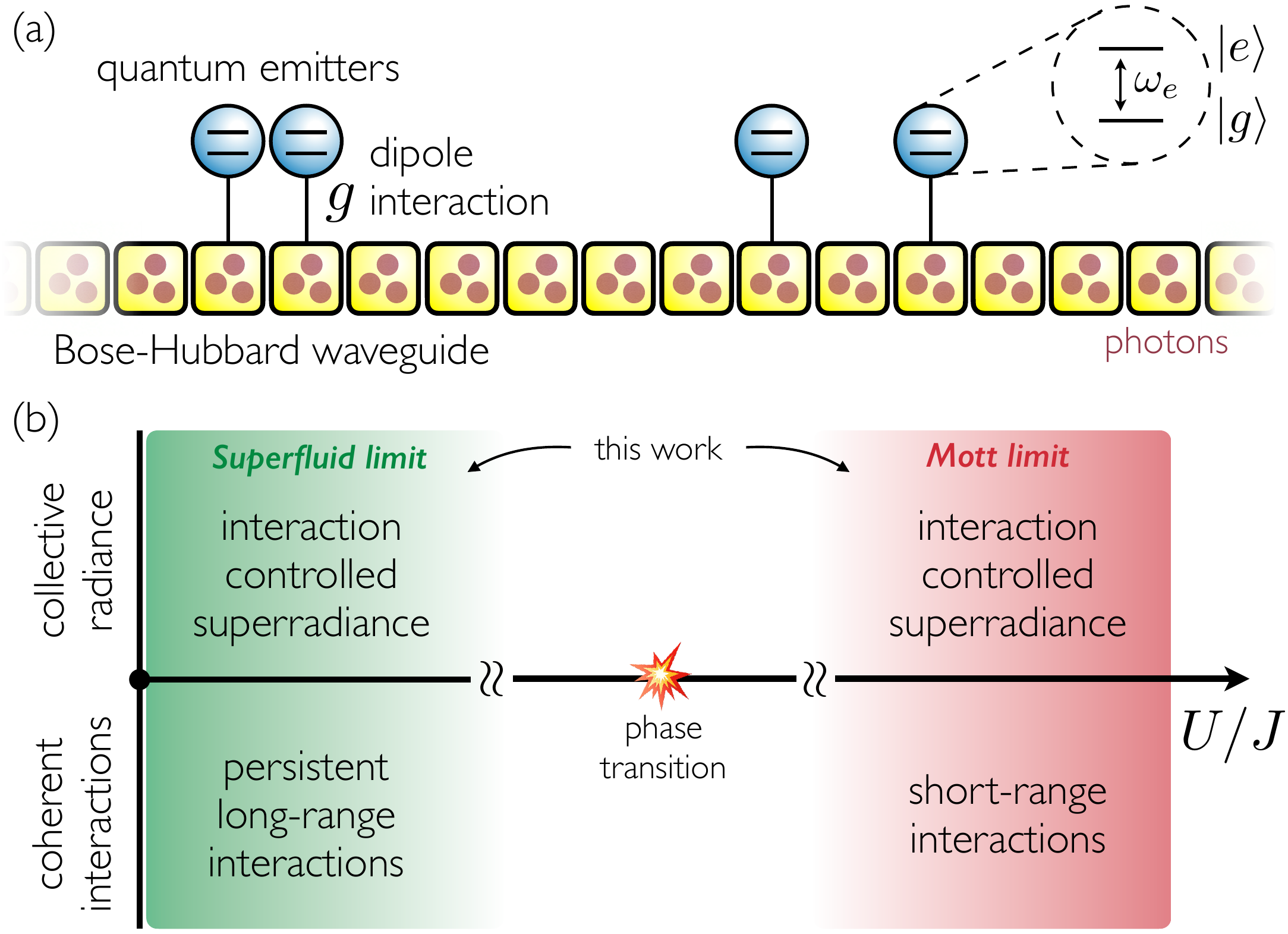}
		\caption{
			{\textit{Setup and summary of main results.} 
				(a) Bose-Hubbard Waveguide QED setup.
				Two-level quantum emitters (blue spheres) are coupled via dipole interaction to a  Bose-Hubbard waveguide (yellow squares array). The internal transition frequency of the emitters' between the excited and ground states $\ket{e}$ and $\ket{g}$ is $\omega_e$. 
				The waveguide array hosts coherent photon hopping at rate $J$ and repulsive local interaction with strength $U$.
				In the limit of many photons (red dots) in the waveguide, two regimes can be identified. 
				In the superfluid (Mott) regime the waveguide can be mapped to Bogoliubov quasiparticle (doublons and holons) living on top of a many-body ``superfluid'' (Mott insulating) ground state. 
				(b) Phase diagram and main results. In the superfluid and Mott limits, photon-photon interactions can trigger a collective superradiant burst. 
				Coherent interactions between quantum emitters become distant-independent in the superfluid limit due to the ground state delocalized nature, while they are short-range in the Mott regime.
			}
		} 
		\label{fig:setup}
	\end{figure}

	\section{Bose-Hubbard waveguide}\label{Sec:BHwg}
	We consider a 1D Bose-Hubbard (BH) waveguide, a tight-binding array of $N_p$ photonic modes with on-site repulsion described by:
	\begin{eqnarray}\label{eq:BHHamilt}
		\hat H_\text{BH} 
		& = & 
		\omega_c \sum_r \hat a_r^\dagger \hat a_r
		- J \sum_r \left( \hat a_r^\dagger \hat a_{r+1} + \text{H.c.} \right) \nonumber\\
		&& + \frac{U}{2} \sum_r \hat a_r^\dagger \hat a_r^\dagger \hat a_r \hat a_r,
	\end{eqnarray}
	where $\hat a_r$ annihilates a photon at site $r$,  the resonator frequency is $\omega_c \gg J,U$~\cite{WangPRL2020,ScigliuzzoPRX2022}, $J>0$ is the hopping rate, and $U>0$ is the repulsive interaction strength (the lattice spacing $d$ is set to 1). Unlike previous studies focusing on attractive interactions~\cite{WangPRL2020,TalukdarPRA2022,WangPRRes2024}, we consider here repulsive photon-photon interactions.
	
	For large photon numbers, the competition between tunneling ($J$) and repulsion ($U$) produces two distinct regimes. In the weak interaction limit ($U\ll J$), the many-body ground state is a photon quasicondensate $\ket{\text{SF}}$, which would become a proper superfluid in 2D~\cite{pitaevskii2016bose}. Conversely, for strong interactions ($J\ll U$), photons localize into a Mott insulator $\ket{\text{MI}}$. While a true quantum phase transition occurs in 2D~\cite{GreinerNature2002}, the 1D system exhibits clear crossover behavior between these limits~\cite{KuhnerPRB1998}.
	
	To treat the waveguide as a photonic bath for emitters (see next Section), we employ effective theories in each limit. For the superfluid phase, we use the Bogoliubov theory of weakly interacting Bose gas~\cite{vanOostenPRA2001,stoof2009ultracold,pitaevskii2016bose}, while the Mott regime is analyzed via the formalism of Barmettler et al.~\cite{BarmettlerPRA2012}. This maps the interacting Hamiltonian to quadratic quasiparticle models, enabling analytic solutions.
	
	\begin{figure}
		\centering
		\includegraphics[width=\columnwidth]{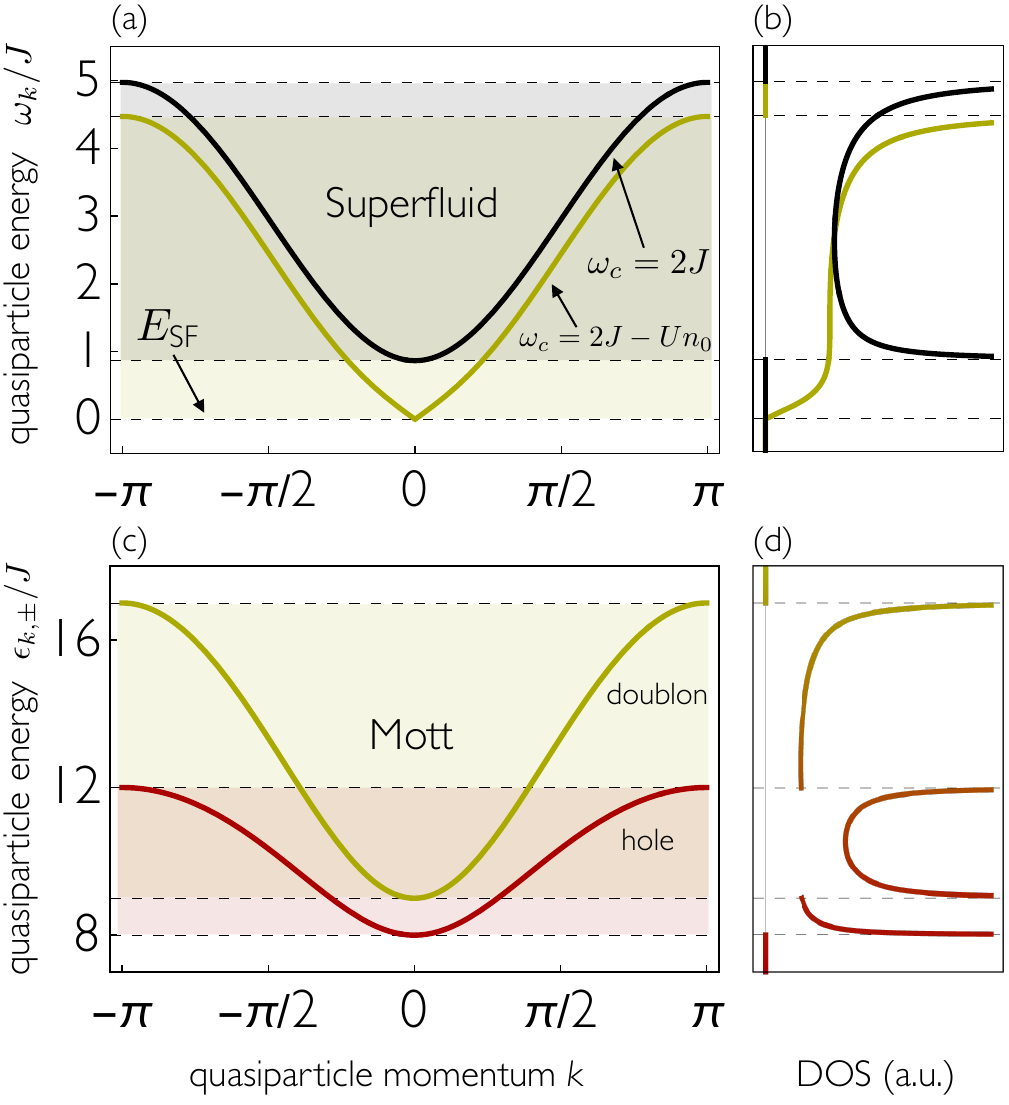}
		\caption{
			\textit{Quasiparticle dispersion and DOS.} 
			(a) Superfluid ($U=J/2$) quasiparticle dispersion for $\omega_c = 2J - U n_0$ (yellow) and $\omega_c =2J$ (black), with corresponding DOS in (b). The DOS, up to a prefactor, is  the single-emitter decay rate $\Gamma^\text{SF}_{ii}(\omega)$. $E_\text{SF}$ marks the ground state energy (zero reference). (c) Mott insulator ($U=3J$) excitations: doublons (yellow) and holons (red). (d) Corresponding DOS, proportional to $\Gamma^\text{MI}_{ii}(\omega)$. The shaded stripes highlight the bandwidth corresponding to each band. Parameters: $n_0=\bar n=1$, $\omega_c=10J$ (c-d), $g=0.1J$.
		} 
		\label{fig:bands}
	\end{figure}
	
	In the superfluid limit ($U \ll J$), we consider the BH waveguide populated by an average number of  $N$ photons, with $N_0$ photons occupying the $k=0$ momentum state such that
	$N = \sum_k \langle \hat a_k^\dagger \hat a_k \rangle = \langle \hat a_0^\dagger \hat a_0 \rangle + \sum_{k\neq0} \langle \hat a_k^\dagger \hat a_k \rangle = N_0 + \sum_{k\neq0} \langle \hat a_k^\dagger \hat a_k \rangle$.
	First, we write
	the Hamiltonian in momentum space, getting
	$\hat H_\text{BH} = \sum_k (\omega_c + \varepsilon_k) \hat a_k^\dagger \hat a_k + (U/2N_p) \sum_{k,k',q,q'} \hat a_{q'}^\dagger \hat a_q^\dagger \hat a_{k'} \hat a_k \delta_{q+q',k+k'},$
	where $\hat a_r = \sum_k \hat a_k e^{ikr}/\sqrt{N_p}$ and $\varepsilon_k = -2J\cos k$. 
	Then,
	following Bogoliubov theory, we assume $N_0 \gg 1$ and replace $\hat a_0$ with $\sqrt{N_0}$. 
	Keeping terms up to order $N_0$ (see~\cite{SupplMat}) we get, after a Bogoliubov rotation,
	the quadratic Hamiltonian:
	\begin{equation}
		\hat H_\text{BH,SF} = E_\text{SF} + \sum_{k\neq0} \omega_k \hat b_k^\dagger \hat b_k,
	\end{equation}
	with ground state energy $E_\text{SF}$ and Bogoliubov excitations $\hat b_k$ satisfying $\hat b_k\ket{\text{SF}} = 0$ for the quasi-condensate state $\ket{\text{SF}} = (\hat a_{k=0}^\dagger)^{N_0}\ket{0}$. The quasiparticle dispersion is
	$\omega_k = \sqrt{f_k^2 - U^2n_0^2}$
	where
	$f_k = \omega_c + 2Un_0 + \varepsilon_k$ and $n_0 = N_0/N_p$ is the condensate fraction. Setting $\omega_c = 2J - Un_0$ recovers the standard BH dispersion~\cite{vanOostenPRA2001}. The Bogoliubov transformation given by
	$\hat b_k = u_k \hat a_k + v_k \hat a_{-k}^\dagger$, 
	where
	$u_k = \cosh \alpha_k$, $v_k = \sinh \alpha_k,$ and $\alpha_k$ is defined by $\tanh(2\alpha_k) = Un_0/f_k$, connects the quasiparticles to the photonic excitations (see~\cite{SupplMat}). Figure~\ref{fig:bands}(a-b) shows the dispersion and corresponding density of states (DOS).
	
	
	In the Mott limit ($J \ll U$), we follow the theory developed by Barmettler et al.~\cite{BarmettlerPRA2012} (and previously in Refs.~\cite{AltmanPRL2002,HuberPRB2007}), noting that our Hamiltonian~\eqref{eq:BHHamilt} differs slightly from the one considered in Refs.~\cite{CheneauNature2012,BarmettlerPRA2012} (see~\cite{SupplMat}). 
	For zero hopping ($J=0$), the ground state is perfectly localized:
	$\ket{\text{MI}} = \bigotimes_r \ket{\bar n}_r$
	where $\bar n $ is the integer filling factor. When small but finite hopping $J$ is introduced the local density fluctuations around the average filling are limited. We can thus restrict the local Hilbert space to three states: $\{\ket{\bar n}_r, \ket{\bar n \pm 1}_r\}$. 
	The original bosonic operators are expressed in terms of auxiliary bosonic operators as:
	$\hat a_r = \sqrt{\bar n + 1}\, \hat b_{r,+} + \sqrt{\bar n} \,\hat b_{r,-}^\dagger$,
	where
	$\hat b_{r,+}^\dagger$ creates an excess particle (doublon), $\hat b_{r,+}^\dagger \ket{\bar n}_r = \ket{\bar n + 1}_r$, and
	$\hat b_{r,-}^\dagger$ creates a hole, $\hat b_{r,-}^\dagger \ket{\bar n}_r = \ket{\bar n - 1}_r$.
	The Fock state $\ket{\bar n}_r$ serves as the vacuum for both auxiliary operators: $\hat b_{r,\pm}\ket{\text{MI}} = 0$. While these auxiliary operators obey bosonic commutation relations, they must satisfy the hardcore boson constraints:
	$\hat b_{r,\pm}^2 = 0$ (no double-creation on one site)
	and
	$\hat b_{r,+}^\dagger \hat b_{r,+} \hat b_{r,-}^\dagger \hat b_{r,-} \equiv \hat n_{r,+} \hat n_{r,-}= 0$ (no simultaneous doublon and holon on one site).
	
	To enforce these constraints, we employ the generalized Jordan-Wigner transformation~\cite{Jordan1928,BatistaPRL2001}, mapping the auxiliary bosons to fermionic operators $\hat c_{r,\pm}$:
	$\hat b_{r,\pm} = \hat Z_{r,\pm} \hat c_{r,\pm}$,
	where $\hat Z_{r,+} = \exp(i\pi \sum_{\sigma=\pm} \sum_{r'<r} \hat n_{r',\sigma})$ and
	$\hat Z_{r,-} = \hat Z_{r,+} \exp(i\pi \hat n_{r,+})$ are nonlocal string operators. This automatically enforces the hardcore constraint.
	The remaining constraint against double occupancies of different species
	is treated by projecting the Hamiltonian with $\mathcal P = \prod_r (1 - \hat n_{r,+}\hat n_{r,-})$ and making the  unconstrained fermions approximation $\mathcal P  \rightarrow 1$ following Ref.~\cite{BarmettlerPRA2012}.
	
	Once the initial Hamiltonian is written in terms of fermionic operators $\hat c_{r,\pm}$, transforming in momentum space $\hat c_{r,\pm} =  \sum_k \hat c_{k,\pm} e^{ikr}/\sqrt{N_p}$ and performing the Bogoliubov rotation  
	$\hat \gamma_{k,\pm}
	=
	(u_k
	\hat c_{k,\pm}
	-
	v_k
	\hat c_{-k,\mp}^\dagger)
	$,
	where $u_k$ ($v_k$) is real (imaginary) and even (odd) in $k$,
	one can recast the BH waveguide Hamiltonian into 
	\begin{eqnarray}
		\hat H_\T{BH,MI}
		& = &
		E_\T{MI}
		+
		\sum_{k,\sigma}
		\epsilon_{k,\sigma}
		\hat \gamma_{k,\sigma}^\dagger
		\hat \gamma_{k,\sigma}\,.
	\end{eqnarray}
	Here, $E_\T{MI}$ is the ground state energy and the $\hat \gamma_{k,\pm}$'s are the (fermionic) Bogoliubov excitations on top of the ground state, so that, in the limit of vanishing hopping strength, the ground state $\ket{\T{MI}}$ is the vacuum of the Bogoliubov excitations $\hat \gamma_{k,\pm}\ket{\T{MI}}=0$.
	The doublon and holon dispersions are given by $\epsilon_{k,\pm} = \pm \delta_k + \eta_k$ where $2\delta_k = E_+(k) - E_-(k)$ and $2\eta_k = \sqrt{[E_+(k) + E_-(k)]^2 + 4 |\Delta(k)|^2}$. Here 
	$E_+(k)
	=
	\omega_c + U \bar n
	-2J (\bar n + 1) \cos k
	$,  
	$E_-(k)
	=
	\omega_c - U (\bar n - 1)
	-2J \bar n \cos k
	$, and
	$
	\Delta(k)
	=
	-2iJ\sqrt{\bar n(\bar n + 1)}\sin k
	$. Finally, the coefficients of the Bogoliubov transformation are given in the Mott limit by
	$u_k
	=
	\cos(\theta_k/2)$
	and 
	$v_k
	=
	i\sin(\theta_k/2)$, where
	$\theta_k$ is defined by $\tan \theta_k = 2i\Delta(k)/[E_+(k) + E_-(k)]$ (see~\cite{SupplMat}).
	In Fig.~\ref{fig:bands}(c-d) we show the doublon/holon dispersion and the corresponding density of states (DOS).
	
	\section{Bose-Hubbard Waveguide QED}\label{Sec:MA}
	Our main goal is to study how two-level quantum emitters interact with a BH waveguide in its superfluid and Mott insulating phases, see Fig.~\ref{fig:setup}. The complete system is described by the Hamiltonian
	$\hat H = \hat H_\text{qe} + \hat H_\text{BH} + \hat V$.
	The emitters' Hamiltonian $\hat H_\text{qe} = \omega_e \sum_i \hat \sigma_i^\dagger \hat \sigma_i$ describes $N$ identical two-level systems with transition frequency $\omega_e$ between states $\ket{g}$ and $\ket{e}$, where $\hat \sigma_i = \ket{g}_i\bra{e}$ is the lowering operator for the $i$-th emitter.
	The interaction term $\hat V$ represents the dipole coupling between emitters and the photonic field,
	\begin{eqnarray}
		\hat V & = & g\sum_i \hat \sigma_i^x \hat x_{r_i} = g\sum_i (\hat \sigma_i + \hat \sigma_i^\dagger)(\hat a_{r_i} + \hat a_{r_i}^\dagger),
		\label{eq:Vint}
	\end{eqnarray}
	where we  keep the counter-rotating terms--the rotating wave approximation is made  in the derivation of the emitters' master equation (see~\cite{SupplMat}).
	The photonic operators $\hat a_r$ take dramatically different forms in each phase of the waveguide in terms of the quasi-excitations:
	\begin{eqnarray}
		\hat a_{r}^{\text{SF}} & = & \sqrt{n_0} + \frac{1}{\sqrt{N_p}}\sum_{k\neq 0}(u_k \hat b_k - v_k \hat b_{-k}^\dagger)e^{ikr}, \label{eq:aSF} \\
		\hat a_{r}^{\text{MI}} & = & \frac{\sqrt{\bar n + 1}}{\sqrt{N_p}} \hat Z_{r,+} \sum_k e^{ikr}(u_k \hat \gamma_{k,+} + v_k \hat \gamma_{-k,-}^\dagger) \nonumber\\
		&& + \frac{\sqrt{\bar n}}{\sqrt{N_p}} \hat Z_{r,-} \sum_k e^{-ikr}(u_k \hat \gamma_{k,-}^\dagger - v_k \hat \gamma_{-k,+}),
	\end{eqnarray}
	with distinct Bogoliubov coefficients $u_k,v_k$ in each phase. 
	We note that previous works studied impurities in BH baths~\cite{CoscoPRA2018,CaleffiNJP2021}, though focusing on density-density interactions.

	The reduced emitter dynamics may be cast in the following form
	\begin{eqnarray}\label{eq:generalME}
		\dot{\tilde{\rho}} & = & -i[\hat H_\text{eff}(\omega_e), \tilde\rho] + \sum_{ij}\Gamma_{ij}(\omega_e)\mathcal{D}[\hat \sigma_j,\hat \sigma_i^\dagger]\tilde\rho,
	\end{eqnarray}
	where $\tilde{\rho}$ is the joint state of the quantum emitters in interaction picture (with respect to $\hat H _\T{qe} + \hat H _\T{BH}$) and $\mathcal{D}[\hat A,\hat B]\rho = \hat A \rho \hat B - \{\hat B \hat A,\rho\}/2$. The effective Hamiltonian 
	\begin{eqnarray}\label{eq:Heff}
		\hat H_\text{eff}(\omega_e) & = & \sum_{ij}[\Delta_{ij}(\omega_e)\hat \sigma_i^\dagger\hat \sigma_j + \Delta_{ij}(-\omega_e)\hat \sigma_i \hat \sigma_j^\dagger]  
	\end{eqnarray}
	describes coherent interactions mediated by the Bogoliubov excitations, while $\Gamma_{ij}(\omega_e)$ gives collective decay rates in the BH waveguide. Both depend on the bath correlation function (see~\cite{SupplMat}):
	\begin{eqnarray}\label{eq:ACF}
		I_{ij}(\omega) & = & g^2\int_0^\infty dt\, e^{i\omega t}\langle \tilde{\hat{x}}_{r_i}(t)\hat{x}_{r_j}\rangle,
	\end{eqnarray}
	where $\tilde{\hat{ x}}_{r}(t)$ is $\hat{ x}_{r}$ [cf.~Eq.~\eqref{eq:Vint}] in interaction picture  and the average is taken over the BH waveguide ground state ($\ket{\text{SF}}$ or $\ket{\text{MI}}$). 
	In fact, we are assuming that at time $t=0$ the ground state of the BH waveguide is prepared, and that at time $t=0^+$ the atom light coupling $\hat V$ is turned on. 
	The exact form of these quantities reveals how many-body effects modify light-matter interactions. 
	Additionally, the emitters can decay into independent non-guided modes of the surrounding electromagnetic field. We include this effect by adding the local  dissipator $\Gamma'\sum_i\mathcal{D}[\hat \sigma_i,\hat \sigma_i^\dagger]\rho$ to the master equation.
	
	\section{Superfluid and Mott Superradiance}\label{subsec:Superradiance}
	\begin{figure}[t!]
		\centering
		\includegraphics[width=\columnwidth]{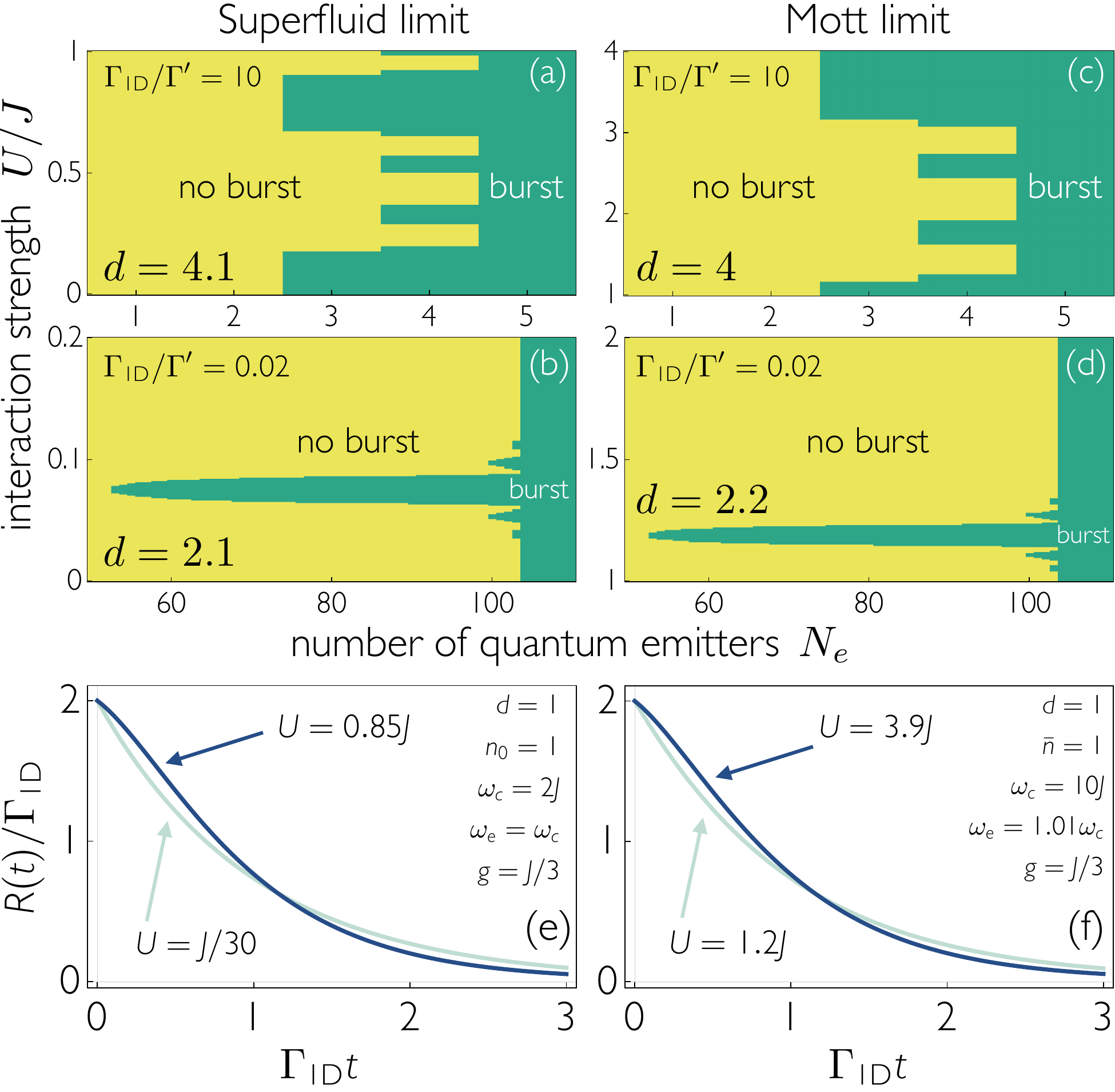}
		\caption{
			{\textit{Superfluid and Mott Superradiance.} 
				Prediction for a superradiant burst in a superfluid (left) or Mott insulating (right) waveguide.  
				The key difference here is that the interatomic distance $d$ and the atomic frequency $\omega_e$ are fixed, still the occurrence of the burst is controlled by the interaction strength $U$, which modulates the momentum $k_\T{1D}$. 
				Superfluid parameters (a-b): $n_0=1$, $\omega_c=2J$, $\omega_e=2J$. 
				Mott parameters (c-d): $\bar n=1$, $\omega_c=10J$,  $\omega_e=11J$. In all panels $g=0.1J$.
				For comparison, see Figure 2 in Ref.~\cite{CardenasPRL2023}.
				Panels (e) and (f): Emission rate $R(t)$, scaled with the single emitter decay rate $\Gamma_{\T{1D}}$, for different values of $U/J$ in the superfluid and Mott limits, respectively. All other parameters are specified within the plot frame.
			}
		} 
		\label{fig:superrad}
	\end{figure}
	We first study the collective decay of $N_e$ excited emitters in our BH waveguide. 
	Superradiance refers to the cooperative enhancement of spontaneous emission due to coherent coupling between the emitters via the bath. This results in a superradiant burst--a short-lived spike in the total emitted power, scaling faster than linearly with the number of emitters~\cite{DickePR1954,GrossPhysicsReports1982}.
	A sufficient condition for the occurrence of a  superradiant burst found in Ref.~\cite{MassonNatCommun2022}
	becomes in the waveguide case~\cite{CardenasPRL2023}:
	\begin{eqnarray}\label{eq:condition}
		\frac{N_e}{2}
		+
		\frac{1}{2N_e}
		\frac{\sin^2 (N_e k_\T{1D} d)}{\sin^2  (k_\T{1D} d)}
		>
		2
		+
		\frac{\Gamma'}{\Gamma_{\T{1D}}}
	\end{eqnarray}
	where we have reintroduced explicitly the interatomic distance $d$ (the lattice constant),  $k_\T{1D}$ is the  momentum selected by the atomic frequency (see below), and $\Gamma_{\T{1D}} = \Gamma_{ii}(\omega_e)$ is the single-emitter decay rate in the BH waveguide.

	We show that, in the superfluid and Mott limits, the photon-photon interaction strength allows to control the occurrence of the superradiant burst. In the standard non-interacting waveguide QED scenario, 
	including a non-negligible parasitic decay,
	the condition~\eqref{eq:condition} can be satisfied (or not) by varying the atomic distances for a fixed atomic frequency $\omega_e$
	(which fixes the momentum $k_{\T{1D}} = \omega_e/v$, $v$ being the speed of light in the waveguide), or vice versa fixing the inter-atomic distance and varying the momentum~\cite{CardenasPRL2023}. 
	
	Including photon-photon interactions with strength $U$, the Bogoliubov momentum becomes dependent on $U$ as well (as the dispersion depends on $U$). This allows then, for a fixed interatomic distance $d$ and fixed atomic frequency $\omega_e$, to tune the  momentum $k_\T{1D}$ so to explore regions of parameter space with and without a burst.
	
	In order to show this, we compute the collective decay rates in both limits which we will label either SF or MI. In the superfluid limit we get (see~\cite{SupplMat})
	\begin{eqnarray}
		\Gamma_{ij}^{\T{SF}}(\omega)
		& = &
		\frac{g^2}{J}
		\mathcal K_\T{SF}(\omega)
		\cos \left[ k_{\T{1D}}^{\T{SF}}|i-j|d \right]
	\end{eqnarray}
	where the prefactor reads $\mathcal K_\T{SF}(\omega) =2J [\Omega- Un_0]/[\Omega\sqrt{4J^2 - (\omega_c + 2 U n_0 - \Omega)^2}]$,  $\Omega=\sqrt{\omega^2+U^2 n_0^2}$ and $k_{\T{1D}}^{\T{SF}}$ is defined by 
	$2J \cos k_{\T{1D}}^{\T{SF}} = \omega_c + 2 U n_0 - \Omega$. The collective decay $\Gamma_{ij}^{\T{SF}}(\omega)$ is non-zero only when the atomic frequency is in resonance with the Bogoliubov band, i.e., $\omega_0<\omega<\omega_\pi$.
	In the non-interacting limit ($U=0$), this reduces to the standard waveguide QED results~\cite{CardenasPRL2023} with the typical band-edge divergences. For a finite interaction strength, the collective decay rate can vanish at the lower band edge   instead~\cite{RecatiPRL2005}.
	In the Mott limit, the collective decay rate is given by
	\begin{equation}\label{eq:colldecMI}
		\Gamma_{ij}^{\T{MI}}(\omega)
		=
		\frac{g^2}{J}
		\sum_{\sigma=\pm}
		\chi_\sigma(\omega)
		\mathcal K_{\T{MI}}^{(\sigma)}(\omega)
		\cos \left[ {k_{\T{1D},\sigma}^\T{MI}}|i-j|d \right]
	\end{equation}
	where the explicit form of $\mathcal K_{\T{MI}}^{(\sigma)}(\omega)$ is more involved than the superfluid one and provided in  Ref.~\cite{SupplMat}. The term 
	$\chi_\sigma(\omega) = \chi_{[\epsilon_{0,\sigma},\epsilon_{\pi,\sigma}]}(\omega)$  is the characteristic function $\chi_A(\omega)$ ($\chi_A(\omega) = 1$ if $\omega\in A$ and 0 otherwise), ensuring that the decay rate is zero if the atomic frequency lies in a doublon/holon bandgap. In Eq.~\eqref{eq:colldecMI}, the  momentum $k_{\T{1D},+}^\T{MI}$ is defined by
	\begin{eqnarray}
		\label{eq:momentumMI}
		\cos {k_{\T{1D},+}^\T{MI}}
		& = & 
		\frac{4J^2 \bar n(\bar n +1) 
		}{2J[(2\bar n +1)\omega_c+U+\omega]}
		\nonumber\\
		& &
		+
		\frac{ [\omega_c-U(\bar n-1)+\omega] 
			[\omega_c-U\bar n-\omega]
		}{2J[(2\bar n +1)\omega_c+U+\omega]},
		\quad
	\end{eqnarray}
	and $k_{\T{1D},-}^\T{MI}$ is obtained from  $k_{\T{1D},+}^\T{MI}$ replacing $\omega$ with $-\omega$.
	Using $k_{\T{1D}}^{\T{SF}}$, $k_{\T{1D},\pm}^{\T{MI}}$ and the condition~\eqref{eq:condition}, in Fig.~\ref{fig:superrad} we show the crossover between burst and no-burst regions, for a fixed interatomic distance, as a function of  number of emitters and interaction strength.
	
	Finally, to  quantify the impact of photon-photon interactions on the dynamics of correlated emission, we compute from Eq.~\eqref{eq:generalME} the time-resolved emission rate \(R(t)=\sum_{ij}\Gamma_{ij}(\omega_e)\,\langle \hat\sigma_i^\dagger \hat\sigma_j\rangle_t\), see Fig.~\ref{fig:superrad}(e-f).  
	We consider as a minimal example two identical emitters initially in \(\ket{\psi_\text{init}}=\ket{ee}\). 
	With \(d\) and \(\omega_e\) fixed, tuning only the photon-photon interaction strength \(U\) qualitatively reshapes \(R(t)\) in both the superfluid and Mott regimes, demonstrating that \(U\) alone controls correlated emission.

	\begin{figure}
		\centering
		\includegraphics[width=\columnwidth]{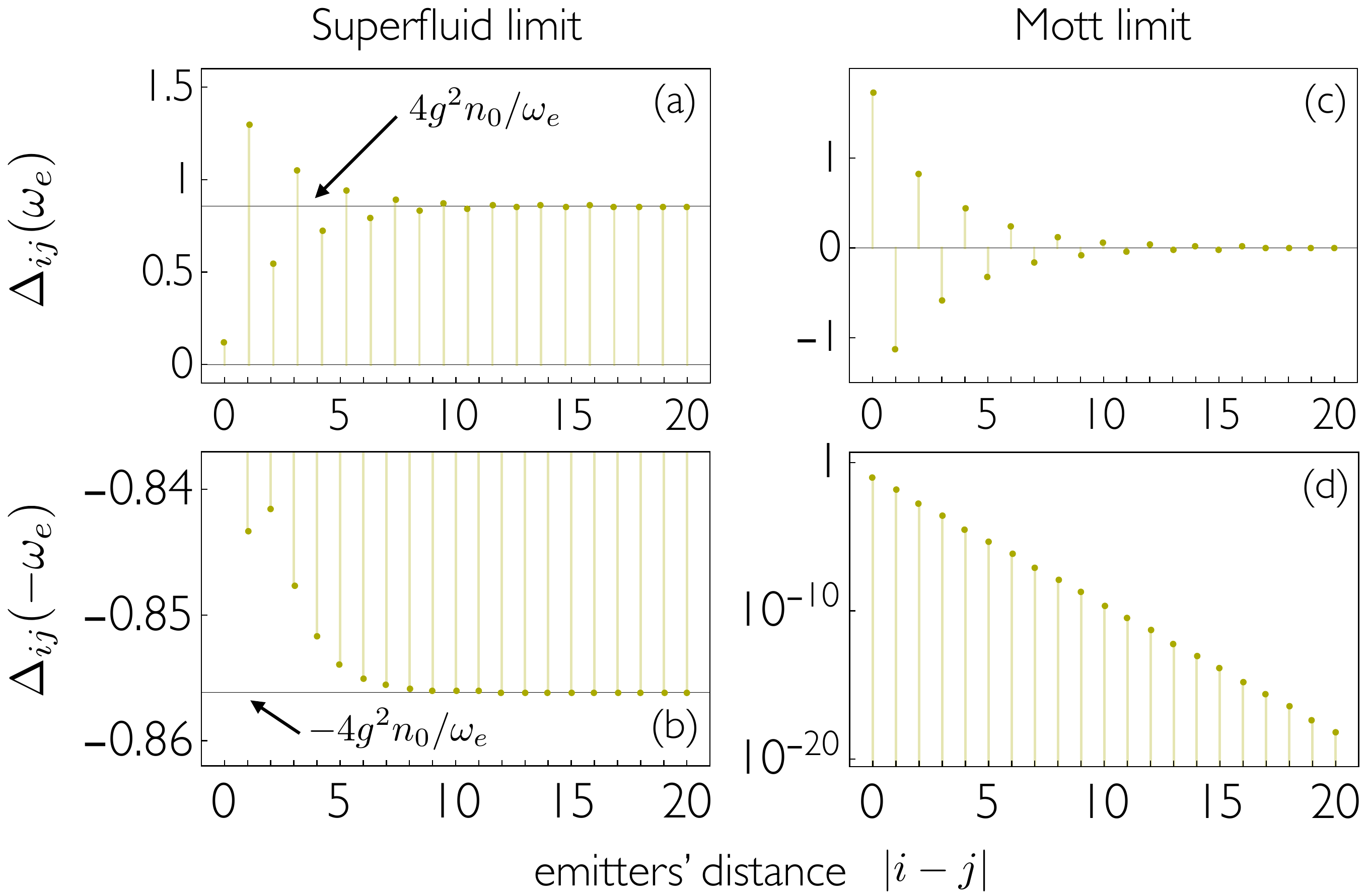}
		\caption{
			{\textit{Bogoliubov-mediated interactions.} 
				Coherent couplings $\Delta_{ij}(\pm\omega_e)$, as in Eq.~\eqref{eq:Heff}, in units of $g^2/J$. In the superfluid limit (a-b), residual  dipole-dipole interactions persist even at large emitters' distance as a result of the underlying many-body ground state driving the emitters. Parameter values in (a-b): $U=J/2$, $n_0=1$, $\omega_c = 2J -U n_0$, $\omega_e = \omega_\pi + 0.2J$. In the Mott limit (c-d), the emitter's interact with the doublon/holon quasiparticles as in the standard waveguide QED case, leading to and exponentially decreasing coupling strength with the emitters' distance.  Parameter values in (c-d): $U=2J$, $\bar n=1$, $\omega_c = 10J$, $\omega_e = E_+(\pi) + 0.2J$. In all panels $g=0.1J$.
			}
		} 
		\label{fig:coherint}
	\end{figure}

	\section{Bogoliubov-mediated  interactions}\label{subsec:interactions}
	We now study the coherent part of the master equation~\eqref{eq:generalME}. For both the superfluid and Mott insulating case we will assume the atomic frequency is off-resonant with the quasiparticle spectrum, namely above it. Therefore, the collective decay rates vanish and the dynamics is fully coherent and driven by $\hat H_\T{eff}$.
	
	The methods we use are formally equivalent to those employed in the studies of photon-mediated interactions. The key difference here is the physical picture. The quantum emitters are coupled to an interacting environment that is in a many-body ground state, and the dipole-dipole interactions are now mediated by the Bogoliubov excitations  on top of the ground state.  
	
	Contrary to the collective decay rates, we approximate the dispersion relations (for weak $U$ and $J$ in the superfluid and Mott limit, respectively).  This allows for the  analytical computation of the coherent couplings $\Delta_{ij}(\omega_e)$ from the autocorrelation function using the residue theorem (see~\cite{SupplMat}).
	
	We find that in the superfluid limit, the coherent interactions have a component that decays exponentially with emitters' distance and one that is independent on it, namely
	\begin{eqnarray}\label{eq:coherSF}
		\Delta_{ij}^{\T{SF}}(\omega_e)
		& = &
		\frac{4 
			g^2
			n_0}{\omega_e }
		+
		\frac{2g^2}{\omega_e}
		F_1(U)
		e^{-|i-j|/\Lambda_1(U)}
		\nonumber\\
		& &
		+
		\frac{2g^2}{\omega_e}
		F_2(U,\omega_e)
		e^{-|i-j|/\Lambda_2(U,\omega_e)}
	\end{eqnarray}
	where the explicit form of $F_1(U)$ and $F_2(U,\omega)$ is not essential here and given in the Supplemental Material~\cite{SupplMat}. Eq.~\eqref{eq:coherSF} is formally valid both for $\omega_e> \omega_\pi$ (upper bandgap) and for $\omega_e<0$ [needed for Eq.~\eqref{eq:Heff}].

	
	Figure~\ref{fig:coherint}(a-b), showing the coherent interactions' strength as a function of emitters' distance [Eq.~\eqref{eq:coherSF}], highlights a central message of this work. In standard (non-interacting) waveguide QED ($U=0$), resonant emitters (in-band) feature all-to-all, long-range dissipative couplings, whereas emitters inside a photonic band gap interact coherently but only over a short range dictated by the exponential tales of atom-photon bound states. In the superfluid case, we find a stark difference: when the emitters lie in a (Bogoliubov) band gap, the interactions remain coherent yet  persisting at long-range.

	In the Mott limit, the Bogoliubov-mediated interactions behave similarly to bandgap photon-mediated interactions in waveguide QED setups, having only the exponential dependence on the emitters' distance. Their explicit expression and derivation is detailed in the Supplemental Material~\cite{SupplMat}. In Fig.~\ref{fig:coherint}, we show the coupling strengths as a function of emitters' distance in the superfluid and Mott limits.

	\section{Conclusion}
	We have shown that many-body photonic environments enable access to previously unexplored regimes of light-matter interaction. By coupling quantum emitters to a Bose-Hubbard waveguide, we demonstrated two key phenomena. 
	First, photon-photon interactions can trigger a superradiant bursts without changing emitter spacing or frequency.
	Second, by breaking the noninteracting short-range/coherence correspondence, the superfluid case furnishes a practical pathway to long-range coherent gates via band-gap engineering. The interaction strength and range are controllable through detuning and interactions, making the effect compatible with existing waveguide-QED platforms and amenable to scalable gate design.

	Our theoretical framework, combining quantum optical and many-body techniques, provides a powerful tool for studying correlated photonic environments. This work opens several research directions, including exploration of topological interacting waveguides, higher-dimensional nonlinear lattices, and nonequilibrium quantum phases. More broadly, it establishes a 
	pathway toward harnessing many-body quantum optical phenomena for quantum simulation and information processing.
	
	\section*{Acknowledgments}
	I thank Ana Asenjo-Garcia, Cosimo C.~Rusconi, Eric Sierra, Fabio Caleffi, Peter Rabl and Bennet Windt for useful discussions.
	I acknowledge financial support by the Fulbright Research Scholar Program and by the European Union-Next Generation EU with the project ``Quantum Optics in Many-Body photonic Environments'' (QOMBE) code SOE2024\_0000084-CUP B77G24000480006.

	\nocite{breuer2002theory}
	\makeatletter\immediate\write\@auxout{\string\citation{apsrev41Control}}\makeatother
	\bibliographystyle{apsrev4-1}	
	\bibliography{references}

\begin{thebibliography}{58}%
\makeatletter
\providecommand \@ifxundefined [1]{%
 \@ifx{#1\undefined}
}%
\providecommand \@ifnum [1]{%
 \ifnum #1\expandafter \@firstoftwo
 \else \expandafter \@secondoftwo
 \fi
}%
\providecommand \@ifx [1]{%
 \ifx #1\expandafter \@firstoftwo
 \else \expandafter \@secondoftwo
 \fi
}%
\providecommand \natexlab [1]{#1}%
\providecommand \enquote  [1]{``#1''}%
\providecommand \bibnamefont  [1]{#1}%
\providecommand \bibfnamefont [1]{#1}%
\providecommand \citenamefont [1]{#1}%
\providecommand \href@noop [0]{\@secondoftwo}%
\providecommand \href [0]{\begingroup \@sanitize@url \@href}%
\providecommand \@href[1]{\@@startlink{#1}\@@href}%
\providecommand \@@href[1]{\endgroup#1\@@endlink}%
\providecommand \@sanitize@url [0]{\catcode `\\12\catcode `\$12\catcode
  `\&12\catcode `\#12\catcode `\^12\catcode `\_12\catcode `\%12\relax}%
\providecommand \@@startlink[1]{}%
\providecommand \@@endlink[0]{}%
\providecommand \url  [0]{\begingroup\@sanitize@url \@url }%
\providecommand \@url [1]{\endgroup\@href {#1}{\urlprefix }}%
\providecommand \urlprefix  [0]{URL }%
\providecommand \Eprint [0]{\href }%
\providecommand \doibase [0]{http://dx.doi.org/}%
\providecommand \selectlanguage [0]{\@gobble}%
\providecommand \bibinfo  [0]{\@secondoftwo}%
\providecommand \bibfield  [0]{\@secondoftwo}%
\providecommand \translation [1]{[#1]}%
\providecommand \BibitemOpen [0]{}%
\providecommand \bibitemStop [0]{}%
\providecommand \bibitemNoStop [0]{.\EOS\space}%
\providecommand \EOS [0]{\spacefactor3000\relax}%
\providecommand \BibitemShut  [1]{\csname bibitem#1\endcsname}%
\let\auto@bib@innerbib\@empty
\bibitem [{\citenamefont {John}\ and\ \citenamefont
  {Wang}(1990)}]{JohnPRL1990}%
  \BibitemOpen
  \bibfield  {author} {\bibinfo {author} {\bibfnamefont {Sajeev}\ \bibnamefont
  {John}}\ and\ \bibinfo {author} {\bibfnamefont {Jian}\ \bibnamefont {Wang}},\
  }\bibfield  {title} {\enquote {\bibinfo {title} {Quantum electrodynamics near
  a photonic band gap: Photon bound states and dressed atoms},}\ }\href
  {\doibase 10.1103/PhysRevLett.64.2418} {\bibfield  {journal} {\bibinfo
  {journal} {Phys. Rev. Lett.}\ }\textbf {\bibinfo {volume} {64}},\ \bibinfo
  {pages} {2418--2421} (\bibinfo {year} {1990})}\BibitemShut {NoStop}%
\bibitem [{\citenamefont {John}\ and\ \citenamefont
  {Quang}(1994)}]{JohnPRA1994}%
  \BibitemOpen
  \bibfield  {author} {\bibinfo {author} {\bibfnamefont {Sajeev}\ \bibnamefont
  {John}}\ and\ \bibinfo {author} {\bibfnamefont {Tran}\ \bibnamefont
  {Quang}},\ }\bibfield  {title} {\enquote {\bibinfo {title} {Spontaneous
  emission near the edge of a photonic band gap},}\ }\href {\doibase
  10.1103/PhysRevA.50.1764} {\bibfield  {journal} {\bibinfo  {journal} {Phys.
  Rev. A}\ }\textbf {\bibinfo {volume} {50}},\ \bibinfo {pages} {1764--1769}
  (\bibinfo {year} {1994})}\BibitemShut {NoStop}%
\bibitem [{\citenamefont {Shen}\ and\ \citenamefont {Fan}(2005)}]{ShenPRL2005}%
  \BibitemOpen
  \bibfield  {author} {\bibinfo {author} {\bibfnamefont {Jung-Tsung}\
  \bibnamefont {Shen}}\ and\ \bibinfo {author} {\bibfnamefont {Shanhui}\
  \bibnamefont {Fan}},\ }\bibfield  {title} {\enquote {\bibinfo {title}
  {Coherent single photon transport in a one-dimensional waveguide coupled with
  superconducting quantum bits},}\ }\href {\doibase
  10.1103/PhysRevLett.95.213001} {\bibfield  {journal} {\bibinfo  {journal}
  {Phys. Rev. Lett.}\ }\textbf {\bibinfo {volume} {95}},\ \bibinfo {pages}
  {213001} (\bibinfo {year} {2005})}\BibitemShut {NoStop}%
\bibitem [{\citenamefont {Shen}\ and\ \citenamefont {Fan}(2009)}]{ShenPRA2009}%
  \BibitemOpen
  \bibfield  {author} {\bibinfo {author} {\bibfnamefont {Jung-Tsung}\
  \bibnamefont {Shen}}\ and\ \bibinfo {author} {\bibfnamefont {Shanhui}\
  \bibnamefont {Fan}},\ }\bibfield  {title} {\enquote {\bibinfo {title} {Theory
  of single-photon transport in a single-mode waveguide. i. coupling to a
  cavity containing a two-level atom},}\ }\href {\doibase
  10.1103/PhysRevA.79.023837} {\bibfield  {journal} {\bibinfo  {journal} {Phys.
  Rev. A}\ }\textbf {\bibinfo {volume} {79}},\ \bibinfo {pages} {023837}
  (\bibinfo {year} {2009})}\BibitemShut {NoStop}%
\bibitem [{\citenamefont {Goban}\ \emph {et~al.}(2015)\citenamefont {Goban},
  \citenamefont {Hung}, \citenamefont {Hood}, \citenamefont {Yu}, \citenamefont
  {Muniz}, \citenamefont {Painter},\ and\ \citenamefont
  {Kimble}}]{GobanPRL2015}%
  \BibitemOpen
  \bibfield  {author} {\bibinfo {author} {\bibfnamefont {A.}~\bibnamefont
  {Goban}}, \bibinfo {author} {\bibfnamefont {C.-L.}\ \bibnamefont {Hung}},
  \bibinfo {author} {\bibfnamefont {J.~D.}\ \bibnamefont {Hood}}, \bibinfo
  {author} {\bibfnamefont {S.-P.}\ \bibnamefont {Yu}}, \bibinfo {author}
  {\bibfnamefont {J.~A.}\ \bibnamefont {Muniz}}, \bibinfo {author}
  {\bibfnamefont {O.}~\bibnamefont {Painter}}, \ and\ \bibinfo {author}
  {\bibfnamefont {H.~J.}\ \bibnamefont {Kimble}},\ }\bibfield  {title}
  {\enquote {\bibinfo {title} {Superradiance for atoms trapped along a photonic
  crystal waveguide},}\ }\href {\doibase 10.1103/PhysRevLett.115.063601}
  {\bibfield  {journal} {\bibinfo  {journal} {Phys. Rev. Lett.}\ }\textbf
  {\bibinfo {volume} {115}},\ \bibinfo {pages} {063601} (\bibinfo {year}
  {2015})}\BibitemShut {NoStop}%
\bibitem [{\citenamefont {Sipahigil}\ \emph {et~al.}(2016)\citenamefont
  {Sipahigil}, \citenamefont {Evans}, \citenamefont {Sukachev}, \citenamefont
  {Burek}, \citenamefont {Borregaard}, \citenamefont {Bhaskar}, \citenamefont
  {Nguyen}, \citenamefont {Pacheco}, \citenamefont {Atikian}, \citenamefont
  {Meuwly}, \citenamefont {Camacho}, \citenamefont {Jelezko}, \citenamefont
  {Bielejec}, \citenamefont {Park}, \citenamefont {Loncar},\ and\ \citenamefont
  {Lukin}}]{SipahigilScience2016}%
  \BibitemOpen
  \bibfield  {author} {\bibinfo {author} {\bibfnamefont {A.}~\bibnamefont
  {Sipahigil}}, \bibinfo {author} {\bibfnamefont {R.~E.}\ \bibnamefont
  {Evans}}, \bibinfo {author} {\bibfnamefont {D.~D.}\ \bibnamefont {Sukachev}},
  \bibinfo {author} {\bibfnamefont {M.~J.}\ \bibnamefont {Burek}}, \bibinfo
  {author} {\bibfnamefont {J.}~\bibnamefont {Borregaard}}, \bibinfo {author}
  {\bibfnamefont {M.~K.}\ \bibnamefont {Bhaskar}}, \bibinfo {author}
  {\bibfnamefont {C.~T.}\ \bibnamefont {Nguyen}}, \bibinfo {author}
  {\bibfnamefont {J.~L.}\ \bibnamefont {Pacheco}}, \bibinfo {author}
  {\bibfnamefont {H.~A.}\ \bibnamefont {Atikian}}, \bibinfo {author}
  {\bibfnamefont {C.}~\bibnamefont {Meuwly}}, \bibinfo {author} {\bibfnamefont
  {R.~M.}\ \bibnamefont {Camacho}}, \bibinfo {author} {\bibfnamefont
  {F.}~\bibnamefont {Jelezko}}, \bibinfo {author} {\bibfnamefont
  {E.}~\bibnamefont {Bielejec}}, \bibinfo {author} {\bibfnamefont
  {H.}~\bibnamefont {Park}}, \bibinfo {author} {\bibfnamefont {M.}~\bibnamefont
  {Loncar}}, \ and\ \bibinfo {author} {\bibfnamefont {M.~D.}\ \bibnamefont
  {Lukin}},\ }\bibfield  {title} {\enquote {\bibinfo {title} {An integrated
  diamond nanophotonics platform for quantum-optical networks},}\ }\href
  {\doibase 10.1126/science.aah6875} {\bibfield  {journal} {\bibinfo  {journal}
  {Science}\ }\textbf {\bibinfo {volume} {354}},\ \bibinfo {pages} {847--850}
  (\bibinfo {year} {2016})}\BibitemShut {NoStop}%
\bibitem [{\citenamefont {Corzo}\ \emph {et~al.}(2019)\citenamefont {Corzo},
  \citenamefont {Raskop}, \citenamefont {Chandra}, \citenamefont {Sheremet},
  \citenamefont {Gouraud},\ and\ \citenamefont {Laurat}}]{CorzoNature2019}%
  \BibitemOpen
  \bibfield  {author} {\bibinfo {author} {\bibfnamefont {Neil~V.}\ \bibnamefont
  {Corzo}}, \bibinfo {author} {\bibfnamefont {J{\'e}r{\'e}my}\ \bibnamefont
  {Raskop}}, \bibinfo {author} {\bibfnamefont {Aveek}\ \bibnamefont {Chandra}},
  \bibinfo {author} {\bibfnamefont {Alexandra~S.}\ \bibnamefont {Sheremet}},
  \bibinfo {author} {\bibfnamefont {Baptiste}\ \bibnamefont {Gouraud}}, \ and\
  \bibinfo {author} {\bibfnamefont {Julien}\ \bibnamefont {Laurat}},\
  }\bibfield  {title} {\enquote {\bibinfo {title} {Waveguide-coupled single
  collective excitation of atomic arrays},}\ }\href {\doibase
  10.1038/s41586-019-0902-3} {\bibfield  {journal} {\bibinfo  {journal}
  {Nature}\ }\textbf {\bibinfo {volume} {566}},\ \bibinfo {pages} {359--362}
  (\bibinfo {year} {2019})}\BibitemShut {NoStop}%
\bibitem [{\citenamefont {Foster}\ \emph {et~al.}(2019)\citenamefont {Foster},
  \citenamefont {Hallett}, \citenamefont {Iorsh}, \citenamefont {Sheldon},
  \citenamefont {Godsland}, \citenamefont {Royall}, \citenamefont {Clarke},
  \citenamefont {Shelykh}, \citenamefont {Fox}, \citenamefont {Skolnick},
  \citenamefont {Itskevich},\ and\ \citenamefont {Wilson}}]{FosterPRL2019}%
  \BibitemOpen
  \bibfield  {author} {\bibinfo {author} {\bibfnamefont {A.~P.}\ \bibnamefont
  {Foster}}, \bibinfo {author} {\bibfnamefont {D.}~\bibnamefont {Hallett}},
  \bibinfo {author} {\bibfnamefont {I.~V.}\ \bibnamefont {Iorsh}}, \bibinfo
  {author} {\bibfnamefont {S.~J.}\ \bibnamefont {Sheldon}}, \bibinfo {author}
  {\bibfnamefont {M.~R.}\ \bibnamefont {Godsland}}, \bibinfo {author}
  {\bibfnamefont {B.}~\bibnamefont {Royall}}, \bibinfo {author} {\bibfnamefont
  {E.}~\bibnamefont {Clarke}}, \bibinfo {author} {\bibfnamefont {I.~A.}\
  \bibnamefont {Shelykh}}, \bibinfo {author} {\bibfnamefont {A.~M.}\
  \bibnamefont {Fox}}, \bibinfo {author} {\bibfnamefont {M.~S.}\ \bibnamefont
  {Skolnick}}, \bibinfo {author} {\bibfnamefont {I.~E.}\ \bibnamefont
  {Itskevich}}, \ and\ \bibinfo {author} {\bibfnamefont {L.~R.}\ \bibnamefont
  {Wilson}},\ }\bibfield  {title} {\enquote {\bibinfo {title} {Tunable photon
  statistics exploiting the fano effect in a waveguide},}\ }\href {\doibase
  10.1103/PhysRevLett.122.173603} {\bibfield  {journal} {\bibinfo  {journal}
  {Phys. Rev. Lett.}\ }\textbf {\bibinfo {volume} {122}},\ \bibinfo {pages}
  {173603} (\bibinfo {year} {2019})}\BibitemShut {NoStop}%
\bibitem [{\citenamefont {Mirhosseini}\ \emph {et~al.}(2019)\citenamefont
  {Mirhosseini}, \citenamefont {Kim}, \citenamefont {Zhang}, \citenamefont
  {Sipahigil}, \citenamefont {Dieterle}, \citenamefont {Keller}, \citenamefont
  {Asenjo-Garcia}, \citenamefont {Chang},\ and\ \citenamefont
  {Painter}}]{MirhosseiniNature2019}%
  \BibitemOpen
  \bibfield  {author} {\bibinfo {author} {\bibfnamefont {Mohammad}\
  \bibnamefont {Mirhosseini}}, \bibinfo {author} {\bibfnamefont {Eunjong}\
  \bibnamefont {Kim}}, \bibinfo {author} {\bibfnamefont {Xueyue}\ \bibnamefont
  {Zhang}}, \bibinfo {author} {\bibfnamefont {Alp}\ \bibnamefont {Sipahigil}},
  \bibinfo {author} {\bibfnamefont {Paul~B.}\ \bibnamefont {Dieterle}},
  \bibinfo {author} {\bibfnamefont {Andrew~J.}\ \bibnamefont {Keller}},
  \bibinfo {author} {\bibfnamefont {Ana}\ \bibnamefont {Asenjo-Garcia}},
  \bibinfo {author} {\bibfnamefont {Darrick~E.}\ \bibnamefont {Chang}}, \ and\
  \bibinfo {author} {\bibfnamefont {Oskar}\ \bibnamefont {Painter}},\
  }\bibfield  {title} {\enquote {\bibinfo {title} {Cavity quantum
  electrodynamics with atom-like mirrors},}\ }\href {\doibase
  10.1038/s41586-019-1196-1} {\bibfield  {journal} {\bibinfo  {journal}
  {Nature}\ }\textbf {\bibinfo {volume} {569}},\ \bibinfo {pages} {692--697}
  (\bibinfo {year} {2019})}\BibitemShut {NoStop}%
\bibitem [{\citenamefont {Kannan}\ \emph {et~al.}(2020)\citenamefont {Kannan},
  \citenamefont {Ruckriegel}, \citenamefont {Campbell}, \citenamefont
  {Frisk~Kockum}, \citenamefont {Braum{\"u}ller}, \citenamefont {Kim},
  \citenamefont {Kjaergaard}, \citenamefont {Krantz}, \citenamefont {Melville},
  \citenamefont {Niedzielski}, \citenamefont {Veps{\"a}l{\"a}inen},
  \citenamefont {Winik}, \citenamefont {Yoder}, \citenamefont {Nori},
  \citenamefont {Orlando}, \citenamefont {Gustavsson},\ and\ \citenamefont
  {Oliver}}]{KannanNature2020}%
  \BibitemOpen
  \bibfield  {author} {\bibinfo {author} {\bibfnamefont {Bharath}\ \bibnamefont
  {Kannan}}, \bibinfo {author} {\bibfnamefont {Max~J.}\ \bibnamefont
  {Ruckriegel}}, \bibinfo {author} {\bibfnamefont {Daniel~L.}\ \bibnamefont
  {Campbell}}, \bibinfo {author} {\bibfnamefont {Anton}\ \bibnamefont
  {Frisk~Kockum}}, \bibinfo {author} {\bibfnamefont {Jochen}\ \bibnamefont
  {Braum{\"u}ller}}, \bibinfo {author} {\bibfnamefont {David~K.}\ \bibnamefont
  {Kim}}, \bibinfo {author} {\bibfnamefont {Morten}\ \bibnamefont
  {Kjaergaard}}, \bibinfo {author} {\bibfnamefont {Philip}\ \bibnamefont
  {Krantz}}, \bibinfo {author} {\bibfnamefont {Alexander}\ \bibnamefont
  {Melville}}, \bibinfo {author} {\bibfnamefont {Bethany~M.}\ \bibnamefont
  {Niedzielski}}, \bibinfo {author} {\bibfnamefont {Antti}\ \bibnamefont
  {Veps{\"a}l{\"a}inen}}, \bibinfo {author} {\bibfnamefont {Roni}\ \bibnamefont
  {Winik}}, \bibinfo {author} {\bibfnamefont {Jonilyn~L.}\ \bibnamefont
  {Yoder}}, \bibinfo {author} {\bibfnamefont {Franco}\ \bibnamefont {Nori}},
  \bibinfo {author} {\bibfnamefont {Terry~P.}\ \bibnamefont {Orlando}},
  \bibinfo {author} {\bibfnamefont {Simon}\ \bibnamefont {Gustavsson}}, \ and\
  \bibinfo {author} {\bibfnamefont {William~D.}\ \bibnamefont {Oliver}},\
  }\bibfield  {title} {\enquote {\bibinfo {title} {Waveguide quantum
  electrodynamics with superconducting artificial giant atoms},}\ }\href
  {\doibase 10.1038/s41586-020-2529-9} {\bibfield  {journal} {\bibinfo
  {journal} {Nature}\ }\textbf {\bibinfo {volume} {583}},\ \bibinfo {pages}
  {775--779} (\bibinfo {year} {2020})}\BibitemShut {NoStop}%
\bibitem [{\citenamefont {Sheremet}\ \emph {et~al.}(2023)\citenamefont
  {Sheremet}, \citenamefont {Petrov}, \citenamefont {Iorsh}, \citenamefont
  {Poshakinskiy},\ and\ \citenamefont {Poddubny}}]{SheremetRMP2023}%
  \BibitemOpen
  \bibfield  {author} {\bibinfo {author} {\bibfnamefont {Alexandra~S.}\
  \bibnamefont {Sheremet}}, \bibinfo {author} {\bibfnamefont {Mihail~I.}\
  \bibnamefont {Petrov}}, \bibinfo {author} {\bibfnamefont {Ivan~V.}\
  \bibnamefont {Iorsh}}, \bibinfo {author} {\bibfnamefont {Alexander~V.}\
  \bibnamefont {Poshakinskiy}}, \ and\ \bibinfo {author} {\bibfnamefont
  {Alexander~N.}\ \bibnamefont {Poddubny}},\ }\bibfield  {title} {\enquote
  {\bibinfo {title} {Waveguide quantum electrodynamics: Collective radiance and
  photon-photon correlations},}\ }\href {\doibase 10.1103/RevModPhys.95.015002}
  {\bibfield  {journal} {\bibinfo  {journal} {Rev. Mod. Phys.}\ }\textbf
  {\bibinfo {volume} {95}},\ \bibinfo {pages} {015002} (\bibinfo {year}
  {2023})}\BibitemShut {NoStop}%
\bibitem [{\citenamefont {Ciccarello}\ \emph {et~al.}(2024)\citenamefont
  {Ciccarello}, \citenamefont {Lodahl},\ and\ \citenamefont
  {Schneble}}]{CiccarelloOptPhotonNews2024}%
  \BibitemOpen
  \bibfield  {author} {\bibinfo {author} {\bibfnamefont {Francesco}\
  \bibnamefont {Ciccarello}}, \bibinfo {author} {\bibfnamefont {Peter}\
  \bibnamefont {Lodahl}}, \ and\ \bibinfo {author} {\bibfnamefont {Dominik}\
  \bibnamefont {Schneble}},\ }\bibfield  {title} {\enquote {\bibinfo {title}
  {Waveguide quantum electrodynamics},}\ }\href {\doibase
  10.1364/OPN.35.5.000034} {\bibfield  {journal} {\bibinfo  {journal} {Opt.
  Photon. News}\ }\textbf {\bibinfo {volume} {35}},\ \bibinfo {pages} {34--41}
  (\bibinfo {year} {2024})}\BibitemShut {NoStop}%
\bibitem [{\citenamefont {Dicke}(1954)}]{DickePR1954}%
  \BibitemOpen
  \bibfield  {author} {\bibinfo {author} {\bibfnamefont {R.~H.}\ \bibnamefont
  {Dicke}},\ }\bibfield  {title} {\enquote {\bibinfo {title} {Coherence in
  spontaneous radiation processes},}\ }\href {\doibase 10.1103/PhysRev.93.99}
  {\bibfield  {journal} {\bibinfo  {journal} {Phys. Rev.}\ }\textbf {\bibinfo
  {volume} {93}},\ \bibinfo {pages} {99--110} (\bibinfo {year}
  {1954})}\BibitemShut {NoStop}%
\bibitem [{\citenamefont {Gross}\ and\ \citenamefont
  {Haroche}(1982)}]{GrossPhysicsReports1982}%
  \BibitemOpen
  \bibfield  {author} {\bibinfo {author} {\bibfnamefont {M.}~\bibnamefont
  {Gross}}\ and\ \bibinfo {author} {\bibfnamefont {S.}~\bibnamefont
  {Haroche}},\ }\bibfield  {title} {\enquote {\bibinfo {title} {Superradiance:
  An essay on the theory of collective spontaneous emission},}\ }\href
  {\doibase https://doi.org/10.1016/0370-1573(82)90102-8} {\bibfield  {journal}
  {\bibinfo  {journal} {Physics Reports}\ }\textbf {\bibinfo {volume} {93}},\
  \bibinfo {pages} {301--396} (\bibinfo {year} {1982})}\BibitemShut {NoStop}%
\bibitem [{\citenamefont {Asenjo-Garcia}\ \emph {et~al.}(2017)\citenamefont
  {Asenjo-Garcia}, \citenamefont {Moreno-Cardoner}, \citenamefont {Albrecht},
  \citenamefont {Kimble},\ and\ \citenamefont {Chang}}]{AsenjoGarciaPRX2017}%
  \BibitemOpen
  \bibfield  {author} {\bibinfo {author} {\bibfnamefont {A.}~\bibnamefont
  {Asenjo-Garcia}}, \bibinfo {author} {\bibfnamefont {M.}~\bibnamefont
  {Moreno-Cardoner}}, \bibinfo {author} {\bibfnamefont {A.}~\bibnamefont
  {Albrecht}}, \bibinfo {author} {\bibfnamefont {H.~J.}\ \bibnamefont
  {Kimble}}, \ and\ \bibinfo {author} {\bibfnamefont {D.~E.}\ \bibnamefont
  {Chang}},\ }\bibfield  {title} {\enquote {\bibinfo {title} {Exponential
  improvement in photon storage fidelities using subradiance and ``selective
  radiance'' in atomic arrays},}\ }\href {\doibase 10.1103/PhysRevX.7.031024}
  {\bibfield  {journal} {\bibinfo  {journal} {Phys. Rev. X}\ }\textbf {\bibinfo
  {volume} {7}},\ \bibinfo {pages} {031024} (\bibinfo {year}
  {2017})}\BibitemShut {NoStop}%
\bibitem [{\citenamefont {Cardenas-Lopez}\ \emph {et~al.}(2023)\citenamefont
  {Cardenas-Lopez}, \citenamefont {Masson}, \citenamefont {Zager},\ and\
  \citenamefont {Asenjo-Garcia}}]{CardenasPRL2023}%
  \BibitemOpen
  \bibfield  {author} {\bibinfo {author} {\bibfnamefont {Silvia}\ \bibnamefont
  {Cardenas-Lopez}}, \bibinfo {author} {\bibfnamefont {Stuart~J.}\ \bibnamefont
  {Masson}}, \bibinfo {author} {\bibfnamefont {Zoe}\ \bibnamefont {Zager}}, \
  and\ \bibinfo {author} {\bibfnamefont {Ana}\ \bibnamefont {Asenjo-Garcia}},\
  }\bibfield  {title} {\enquote {\bibinfo {title} {Many-body superradiance and
  dynamical mirror symmetry breaking in waveguide qed},}\ }\href {\doibase
  10.1103/PhysRevLett.131.033605} {\bibfield  {journal} {\bibinfo  {journal}
  {Phys. Rev. Lett.}\ }\textbf {\bibinfo {volume} {131}},\ \bibinfo {pages}
  {033605} (\bibinfo {year} {2023})}\BibitemShut {NoStop}%
\bibitem [{\citenamefont {Albrecht}\ \emph {et~al.}(2019)\citenamefont
  {Albrecht}, \citenamefont {Henriet}, \citenamefont {Asenjo-Garcia},
  \citenamefont {Dieterle}, \citenamefont {Painter},\ and\ \citenamefont
  {Chang}}]{AlbrechtNJP2019}%
  \BibitemOpen
  \bibfield  {author} {\bibinfo {author} {\bibfnamefont {Andreas}\ \bibnamefont
  {Albrecht}}, \bibinfo {author} {\bibfnamefont {Loïc}\ \bibnamefont
  {Henriet}}, \bibinfo {author} {\bibfnamefont {Ana}\ \bibnamefont
  {Asenjo-Garcia}}, \bibinfo {author} {\bibfnamefont {Paul~B}\ \bibnamefont
  {Dieterle}}, \bibinfo {author} {\bibfnamefont {Oskar}\ \bibnamefont
  {Painter}}, \ and\ \bibinfo {author} {\bibfnamefont {Darrick~E}\ \bibnamefont
  {Chang}},\ }\bibfield  {title} {\enquote {\bibinfo {title} {Subradiant states
  of quantum bits coupled to a one-dimensional waveguide},}\ }\href {\doibase
  10.1088/1367-2630/ab0134} {\bibfield  {journal} {\bibinfo  {journal} {New
  Journal of Physics}\ }\textbf {\bibinfo {volume} {21}},\ \bibinfo {pages}
  {025003} (\bibinfo {year} {2019})}\BibitemShut {NoStop}%
\bibitem [{\citenamefont {Gonz\'alez-Tudela}\ and\ \citenamefont
  {Cirac}(2017)}]{TudelaPRL2017}%
  \BibitemOpen
  \bibfield  {author} {\bibinfo {author} {\bibfnamefont {A.}~\bibnamefont
  {Gonz\'alez-Tudela}}\ and\ \bibinfo {author} {\bibfnamefont {J.~I.}\
  \bibnamefont {Cirac}},\ }\bibfield  {title} {\enquote {\bibinfo {title}
  {Quantum emitters in two-dimensional structured reservoirs in the
  nonperturbative regime},}\ }\href {\doibase 10.1103/PhysRevLett.119.143602}
  {\bibfield  {journal} {\bibinfo  {journal} {Phys. Rev. Lett.}\ }\textbf
  {\bibinfo {volume} {119}},\ \bibinfo {pages} {143602} (\bibinfo {year}
  {2017})}\BibitemShut {NoStop}%
\bibitem [{\citenamefont {Gonz{\'{a}}lez-Tudela}\ and\ \citenamefont
  {Cirac}(2017)}]{Tudela2017b}%
  \BibitemOpen
  \bibfield  {author} {\bibinfo {author} {\bibfnamefont {A.}~\bibnamefont
  {Gonz{\'{a}}lez-Tudela}}\ and\ \bibinfo {author} {\bibfnamefont {J.~I.}\
  \bibnamefont {Cirac}},\ }\bibfield  {title} {\enquote {\bibinfo {title}
  {{Markovian and non-Markovian dynamics of quantum emitters coupled to
  two-dimensional structured reservoirs}},}\ }\href {\doibase
  10.1103/PhysRevA.96.043811} {\bibfield  {journal} {\bibinfo  {journal}
  {Physical Review A}\ }\textbf {\bibinfo {volume} {96}},\ \bibinfo {pages}
  {043811} (\bibinfo {year} {2017})}\BibitemShut {NoStop}%
\bibitem [{\citenamefont {Shi}\ \emph {et~al.}(2018)\citenamefont {Shi},
  \citenamefont {Wu}, \citenamefont {Gonz\'alez-Tudela},\ and\ \citenamefont
  {Cirac}}]{ShiNJP2018}%
  \BibitemOpen
  \bibfield  {author} {\bibinfo {author} {\bibfnamefont {T}~\bibnamefont
  {Shi}}, \bibinfo {author} {\bibfnamefont {Y-H}\ \bibnamefont {Wu}}, \bibinfo
  {author} {\bibfnamefont {A}~\bibnamefont {Gonz\'alez-Tudela}}, \ and\
  \bibinfo {author} {\bibfnamefont {J~I}\ \bibnamefont {Cirac}},\ }\bibfield
  {title} {\enquote {\bibinfo {title} {Effective many-body hamiltonians of
  qubit-photon bound states},}\ }\href {\doibase 10.1088/1367-2630/aae4a9}
  {\bibfield  {journal} {\bibinfo  {journal} {New Journal of Physics}\ }\textbf
  {\bibinfo {volume} {20}},\ \bibinfo {pages} {105005} (\bibinfo {year}
  {2018})}\BibitemShut {NoStop}%
\bibitem [{\citenamefont {Bello}\ \emph {et~al.}(2019)\citenamefont {Bello},
  \citenamefont {Platero}, \citenamefont {Cirac},\ and\ \citenamefont
  {González-Tudela}}]{BelloSciAdv2019}%
  \BibitemOpen
  \bibfield  {author} {\bibinfo {author} {\bibfnamefont {M.}~\bibnamefont
  {Bello}}, \bibinfo {author} {\bibfnamefont {G.}~\bibnamefont {Platero}},
  \bibinfo {author} {\bibfnamefont {J.~I.}\ \bibnamefont {Cirac}}, \ and\
  \bibinfo {author} {\bibfnamefont {A.}~\bibnamefont {González-Tudela}},\
  }\bibfield  {title} {\enquote {\bibinfo {title} {Unconventional quantum
  optics in topological waveguide qed},}\ }\href {\doibase
  10.1126/sciadv.aaw0297} {\bibfield  {journal} {\bibinfo  {journal} {Science
  Advances}\ }\textbf {\bibinfo {volume} {5}},\ \bibinfo {pages} {eaaw0297}
  (\bibinfo {year} {2019})}\BibitemShut {NoStop}%
\bibitem [{\citenamefont {S\'anchez-Burillo}\ \emph {et~al.}(2020)\citenamefont
  {S\'anchez-Burillo}, \citenamefont {Porras},\ and\ \citenamefont
  {Gonz\'alez-Tudela}}]{SanchezPRA20}%
  \BibitemOpen
  \bibfield  {author} {\bibinfo {author} {\bibfnamefont {Eduardo}\ \bibnamefont
  {S\'anchez-Burillo}}, \bibinfo {author} {\bibfnamefont {Diego}\ \bibnamefont
  {Porras}}, \ and\ \bibinfo {author} {\bibfnamefont {Alejandro}\ \bibnamefont
  {Gonz\'alez-Tudela}},\ }\bibfield  {title} {\enquote {\bibinfo {title}
  {Limits of photon-mediated interactions in one-dimensional photonic baths},}\
  }\href {\doibase 10.1103/PhysRevA.102.013709} {\bibfield  {journal} {\bibinfo
   {journal} {Phys. Rev. A}\ }\textbf {\bibinfo {volume} {102}},\ \bibinfo
  {pages} {013709} (\bibinfo {year} {2020})}\BibitemShut {NoStop}%
\bibitem [{\citenamefont {Kim}\ \emph {et~al.}(2021)\citenamefont {Kim},
  \citenamefont {Zhang}, \citenamefont {Ferreira}, \citenamefont {Banker},
  \citenamefont {Iverson}, \citenamefont {Sipahigil}, \citenamefont {Bello},
  \citenamefont {Gonz\'alez-Tudela}, \citenamefont {Mirhosseini},\ and\
  \citenamefont {Painter}}]{KimPRX2021}%
  \BibitemOpen
  \bibfield  {author} {\bibinfo {author} {\bibfnamefont {Eunjong}\ \bibnamefont
  {Kim}}, \bibinfo {author} {\bibfnamefont {Xueyue}\ \bibnamefont {Zhang}},
  \bibinfo {author} {\bibfnamefont {Vinicius~S.}\ \bibnamefont {Ferreira}},
  \bibinfo {author} {\bibfnamefont {Jash}\ \bibnamefont {Banker}}, \bibinfo
  {author} {\bibfnamefont {Joseph~K.}\ \bibnamefont {Iverson}}, \bibinfo
  {author} {\bibfnamefont {Alp}\ \bibnamefont {Sipahigil}}, \bibinfo {author}
  {\bibfnamefont {Miguel}\ \bibnamefont {Bello}}, \bibinfo {author}
  {\bibfnamefont {Alejandro}\ \bibnamefont {Gonz\'alez-Tudela}}, \bibinfo
  {author} {\bibfnamefont {Mohammad}\ \bibnamefont {Mirhosseini}}, \ and\
  \bibinfo {author} {\bibfnamefont {Oskar}\ \bibnamefont {Painter}},\
  }\bibfield  {title} {\enquote {\bibinfo {title} {Quantum electrodynamics in a
  topological waveguide},}\ }\href {\doibase 10.1103/PhysRevX.11.011015}
  {\bibfield  {journal} {\bibinfo  {journal} {Phys. Rev. X}\ }\textbf {\bibinfo
  {volume} {11}},\ \bibinfo {pages} {011015} (\bibinfo {year}
  {2021})}\BibitemShut {NoStop}%
\bibitem [{\citenamefont {Scigliuzzo}\ \emph {et~al.}(2022)\citenamefont
  {Scigliuzzo}, \citenamefont {Calaj\`o}, \citenamefont {Ciccarello},
  \citenamefont {Perez~Lozano}, \citenamefont {Bengtsson}, \citenamefont
  {Scarlino}, \citenamefont {Wallraff}, \citenamefont {Chang}, \citenamefont
  {Delsing},\ and\ \citenamefont {Gasparinetti}}]{ScigliuzzoPRX2022}%
  \BibitemOpen
  \bibfield  {author} {\bibinfo {author} {\bibfnamefont {Marco}\ \bibnamefont
  {Scigliuzzo}}, \bibinfo {author} {\bibfnamefont {Giuseppe}\ \bibnamefont
  {Calaj\`o}}, \bibinfo {author} {\bibfnamefont {Francesco}\ \bibnamefont
  {Ciccarello}}, \bibinfo {author} {\bibfnamefont {Daniel}\ \bibnamefont
  {Perez~Lozano}}, \bibinfo {author} {\bibfnamefont {Andreas}\ \bibnamefont
  {Bengtsson}}, \bibinfo {author} {\bibfnamefont {Pasquale}\ \bibnamefont
  {Scarlino}}, \bibinfo {author} {\bibfnamefont {Andreas}\ \bibnamefont
  {Wallraff}}, \bibinfo {author} {\bibfnamefont {Darrick}\ \bibnamefont
  {Chang}}, \bibinfo {author} {\bibfnamefont {Per}\ \bibnamefont {Delsing}}, \
  and\ \bibinfo {author} {\bibfnamefont {Simone}\ \bibnamefont
  {Gasparinetti}},\ }\bibfield  {title} {\enquote {\bibinfo {title}
  {Controlling atom-photon bound states in an array of josephson-junction
  resonators},}\ }\href {\doibase 10.1103/PhysRevX.12.031036} {\bibfield
  {journal} {\bibinfo  {journal} {Phys. Rev. X}\ }\textbf {\bibinfo {volume}
  {12}},\ \bibinfo {pages} {031036} (\bibinfo {year} {2022})}\BibitemShut
  {NoStop}%
\bibitem [{\citenamefont {Leonforte}\ \emph {et~al.}(2024)\citenamefont
  {Leonforte}, \citenamefont {Sun}, \citenamefont {Valenti}, \citenamefont
  {Spagnolo}, \citenamefont {Illuminati}, \citenamefont {Carollo},\ and\
  \citenamefont {Ciccarello}}]{LeonforteQST2025}%
  \BibitemOpen
  \bibfield  {author} {\bibinfo {author} {\bibfnamefont {Luca}\ \bibnamefont
  {Leonforte}}, \bibinfo {author} {\bibfnamefont {Xuejian}\ \bibnamefont
  {Sun}}, \bibinfo {author} {\bibfnamefont {Davide}\ \bibnamefont {Valenti}},
  \bibinfo {author} {\bibfnamefont {Bernardo}\ \bibnamefont {Spagnolo}},
  \bibinfo {author} {\bibfnamefont {Fabrizio}\ \bibnamefont {Illuminati}},
  \bibinfo {author} {\bibfnamefont {Angelo}\ \bibnamefont {Carollo}}, \ and\
  \bibinfo {author} {\bibfnamefont {Francesco}\ \bibnamefont {Ciccarello}},\
  }\bibfield  {title} {\enquote {\bibinfo {title} {Quantum optics with giant
  atoms in a structured photonic bath},}\ }\href {\doibase
  10.1088/2058-9565/ada08d} {\bibfield  {journal} {\bibinfo  {journal} {Quantum
  Science and Technology}\ }\textbf {\bibinfo {volume} {10}},\ \bibinfo {pages}
  {015057} (\bibinfo {year} {2024})}\BibitemShut {NoStop}%
\bibitem [{\citenamefont {Benedetto}\ \emph {et~al.}(2025)\citenamefont
  {Benedetto}, \citenamefont {Gonzalez-Tudela},\ and\ \citenamefont
  {Ciccarello}}]{Benedetto2025dipoledipole}%
  \BibitemOpen
  \bibfield  {author} {\bibinfo {author} {\bibfnamefont {Enrico~Di}\
  \bibnamefont {Benedetto}}, \bibinfo {author} {\bibfnamefont {Alejandro}\
  \bibnamefont {Gonzalez-Tudela}}, \ and\ \bibinfo {author} {\bibfnamefont
  {Francesco}\ \bibnamefont {Ciccarello}},\ }\bibfield  {title} {\enquote
  {\bibinfo {title} {Dipole-dipole interactions mediated by a photonic flat
  band},}\ }\href {\doibase 10.22331/q-2025-03-25-1671} {\bibfield  {journal}
  {\bibinfo  {journal} {{Quantum}}\ }\textbf {\bibinfo {volume} {9}},\ \bibinfo
  {pages} {1671} (\bibinfo {year} {2025})}\BibitemShut {NoStop}%
\bibitem [{\citenamefont {Wang}\ \emph {et~al.}(2024)\citenamefont {Wang},
  \citenamefont {Li}, \citenamefont {Liu}, \citenamefont {Miranowicz},\ and\
  \citenamefont {Nori}}]{WangPRRes2024}%
  \BibitemOpen
  \bibfield  {author} {\bibinfo {author} {\bibfnamefont {Xin}\ \bibnamefont
  {Wang}}, \bibinfo {author} {\bibfnamefont {Jia-Qi}\ \bibnamefont {Li}},
  \bibinfo {author} {\bibfnamefont {Tao}\ \bibnamefont {Liu}}, \bibinfo
  {author} {\bibfnamefont {Adam}\ \bibnamefont {Miranowicz}}, \ and\ \bibinfo
  {author} {\bibfnamefont {Franco}\ \bibnamefont {Nori}},\ }\bibfield  {title}
  {\enquote {\bibinfo {title} {Long-range four-body interactions in structured
  nonlinear photonic waveguides},}\ }\href {\doibase
  10.1103/PhysRevResearch.6.043226} {\bibfield  {journal} {\bibinfo  {journal}
  {Phys. Rev. Res.}\ }\textbf {\bibinfo {volume} {6}},\ \bibinfo {pages}
  {043226} (\bibinfo {year} {2024})}\BibitemShut {NoStop}%
\bibitem [{\citenamefont {Wang}\ \emph {et~al.}(2020)\citenamefont {Wang},
  \citenamefont {Jaako}, \citenamefont {Kirton},\ and\ \citenamefont
  {Rabl}}]{WangPRL2020}%
  \BibitemOpen
  \bibfield  {author} {\bibinfo {author} {\bibfnamefont {Zhihai}\ \bibnamefont
  {Wang}}, \bibinfo {author} {\bibfnamefont {Tuomas}\ \bibnamefont {Jaako}},
  \bibinfo {author} {\bibfnamefont {Peter}\ \bibnamefont {Kirton}}, \ and\
  \bibinfo {author} {\bibfnamefont {Peter}\ \bibnamefont {Rabl}},\ }\bibfield
  {title} {\enquote {\bibinfo {title} {Supercorrelated radiance in nonlinear
  photonic waveguides},}\ }\href {\doibase 10.1103/PhysRevLett.124.213601}
  {\bibfield  {journal} {\bibinfo  {journal} {Phys. Rev. Lett.}\ }\textbf
  {\bibinfo {volume} {124}},\ \bibinfo {pages} {213601} (\bibinfo {year}
  {2020})}\BibitemShut {NoStop}%
\bibitem [{\citenamefont {Kapit}\ \emph {et~al.}(2014)\citenamefont {Kapit},
  \citenamefont {Hafezi},\ and\ \citenamefont {Simon}}]{KapitPRX2014}%
  \BibitemOpen
  \bibfield  {author} {\bibinfo {author} {\bibfnamefont {Eliot}\ \bibnamefont
  {Kapit}}, \bibinfo {author} {\bibfnamefont {Mohammad}\ \bibnamefont
  {Hafezi}}, \ and\ \bibinfo {author} {\bibfnamefont {Steven~H.}\ \bibnamefont
  {Simon}},\ }\bibfield  {title} {\enquote {\bibinfo {title} {Induced
  self-stabilization in fractional quantum hall states of light},}\ }\href
  {\doibase 10.1103/PhysRevX.4.031039} {\bibfield  {journal} {\bibinfo
  {journal} {Phys. Rev. X}\ }\textbf {\bibinfo {volume} {4}},\ \bibinfo {pages}
  {031039} (\bibinfo {year} {2014})}\BibitemShut {NoStop}%
\bibitem [{\citenamefont {Roushan}\ \emph {et~al.}(2017)\citenamefont
  {Roushan}, \citenamefont {Neill}, \citenamefont {Tangpanitanon},
  \citenamefont {Bastidas}, \citenamefont {Megrant}, \citenamefont {Barends},
  \citenamefont {Chen}, \citenamefont {Chen}, \citenamefont {Chiaro},
  \citenamefont {Dunsworth}, \citenamefont {Fowler}, \citenamefont {Foxen},
  \citenamefont {Giustina}, \citenamefont {Jeffrey}, \citenamefont {Kelly},
  \citenamefont {Lucero}, \citenamefont {Mutus}, \citenamefont {Neeley},
  \citenamefont {Quintana}, \citenamefont {Sank}, \citenamefont {Vainsencher},
  \citenamefont {Wenner}, \citenamefont {White}, \citenamefont {Neven},
  \citenamefont {Angelakis},\ and\ \citenamefont
  {Martinis}}]{RoushanScience2017}%
  \BibitemOpen
  \bibfield  {author} {\bibinfo {author} {\bibfnamefont {P.}~\bibnamefont
  {Roushan}}, \bibinfo {author} {\bibfnamefont {C.}~\bibnamefont {Neill}},
  \bibinfo {author} {\bibfnamefont {J.}~\bibnamefont {Tangpanitanon}}, \bibinfo
  {author} {\bibfnamefont {V.~M.}\ \bibnamefont {Bastidas}}, \bibinfo {author}
  {\bibfnamefont {A.}~\bibnamefont {Megrant}}, \bibinfo {author} {\bibfnamefont
  {R.}~\bibnamefont {Barends}}, \bibinfo {author} {\bibfnamefont
  {Y.}~\bibnamefont {Chen}}, \bibinfo {author} {\bibfnamefont {Z.}~\bibnamefont
  {Chen}}, \bibinfo {author} {\bibfnamefont {B.}~\bibnamefont {Chiaro}},
  \bibinfo {author} {\bibfnamefont {A.}~\bibnamefont {Dunsworth}}, \bibinfo
  {author} {\bibfnamefont {A.}~\bibnamefont {Fowler}}, \bibinfo {author}
  {\bibfnamefont {B.}~\bibnamefont {Foxen}}, \bibinfo {author} {\bibfnamefont
  {M.}~\bibnamefont {Giustina}}, \bibinfo {author} {\bibfnamefont
  {E.}~\bibnamefont {Jeffrey}}, \bibinfo {author} {\bibfnamefont
  {J.}~\bibnamefont {Kelly}}, \bibinfo {author} {\bibfnamefont
  {E.}~\bibnamefont {Lucero}}, \bibinfo {author} {\bibfnamefont
  {J.}~\bibnamefont {Mutus}}, \bibinfo {author} {\bibfnamefont
  {M.}~\bibnamefont {Neeley}}, \bibinfo {author} {\bibfnamefont
  {C.}~\bibnamefont {Quintana}}, \bibinfo {author} {\bibfnamefont
  {D.}~\bibnamefont {Sank}}, \bibinfo {author} {\bibfnamefont {A.}~\bibnamefont
  {Vainsencher}}, \bibinfo {author} {\bibfnamefont {J.}~\bibnamefont {Wenner}},
  \bibinfo {author} {\bibfnamefont {T.}~\bibnamefont {White}}, \bibinfo
  {author} {\bibfnamefont {H.}~\bibnamefont {Neven}}, \bibinfo {author}
  {\bibfnamefont {D.~G.}\ \bibnamefont {Angelakis}}, \ and\ \bibinfo {author}
  {\bibfnamefont {J.}~\bibnamefont {Martinis}},\ }\bibfield  {title} {\enquote
  {\bibinfo {title} {Spectroscopic signatures of localization with interacting
  photons in superconducting qubits},}\ }\href {\doibase
  10.1126/science.aao1401} {\bibfield  {journal} {\bibinfo  {journal}
  {Science}\ }\textbf {\bibinfo {volume} {358}},\ \bibinfo {pages} {1175--1179}
  (\bibinfo {year} {2017})}\BibitemShut {NoStop}%
\bibitem [{\citenamefont {Ma}\ \emph {et~al.}(2019)\citenamefont {Ma},
  \citenamefont {Saxberg}, \citenamefont {Owens}, \citenamefont {Leung},
  \citenamefont {Lu}, \citenamefont {Simon},\ and\ \citenamefont
  {Schuster}}]{MaNature2019}%
  \BibitemOpen
  \bibfield  {author} {\bibinfo {author} {\bibfnamefont {Ruichao}\ \bibnamefont
  {Ma}}, \bibinfo {author} {\bibfnamefont {Brendan}\ \bibnamefont {Saxberg}},
  \bibinfo {author} {\bibfnamefont {Clai}\ \bibnamefont {Owens}}, \bibinfo
  {author} {\bibfnamefont {Nelson}\ \bibnamefont {Leung}}, \bibinfo {author}
  {\bibfnamefont {Yao}\ \bibnamefont {Lu}}, \bibinfo {author} {\bibfnamefont
  {Jonathan}\ \bibnamefont {Simon}}, \ and\ \bibinfo {author} {\bibfnamefont
  {David~I.}\ \bibnamefont {Schuster}},\ }\bibfield  {title} {\enquote
  {\bibinfo {title} {A dissipatively stabilized mott insulator of photons},}\
  }\href {\doibase 10.1038/s41586-019-0897-9} {\bibfield  {journal} {\bibinfo
  {journal} {Nature}\ }\textbf {\bibinfo {volume} {566}},\ \bibinfo {pages}
  {51--57} (\bibinfo {year} {2019})}\BibitemShut {NoStop}%
\bibitem [{\citenamefont {Carusotto}\ \emph {et~al.}(2020)\citenamefont
  {Carusotto}, \citenamefont {Houck}, \citenamefont {Koll{\'a}r}, \citenamefont
  {Roushan}, \citenamefont {Schuster},\ and\ \citenamefont
  {Simon}}]{CarusottoNatPhys2020}%
  \BibitemOpen
  \bibfield  {author} {\bibinfo {author} {\bibfnamefont {Iacopo}\ \bibnamefont
  {Carusotto}}, \bibinfo {author} {\bibfnamefont {Andrew~A.}\ \bibnamefont
  {Houck}}, \bibinfo {author} {\bibfnamefont {Alicia~J.}\ \bibnamefont
  {Koll{\'a}r}}, \bibinfo {author} {\bibfnamefont {Pedram}\ \bibnamefont
  {Roushan}}, \bibinfo {author} {\bibfnamefont {David~I.}\ \bibnamefont
  {Schuster}}, \ and\ \bibinfo {author} {\bibfnamefont {Jonathan}\ \bibnamefont
  {Simon}},\ }\bibfield  {title} {\enquote {\bibinfo {title} {Photonic
  materials in circuit quantum electrodynamics},}\ }\href {\doibase
  10.1038/s41567-020-0815-y} {\bibfield  {journal} {\bibinfo  {journal} {Nature
  Physics}\ }\textbf {\bibinfo {volume} {16}},\ \bibinfo {pages} {268--279}
  (\bibinfo {year} {2020})}\BibitemShut {NoStop}%
\bibitem [{\citenamefont {Fisher}\ \emph {et~al.}(1989)\citenamefont {Fisher},
  \citenamefont {Weichman}, \citenamefont {Grinstein},\ and\ \citenamefont
  {Fisher}}]{FisherPRB1989}%
  \BibitemOpen
  \bibfield  {author} {\bibinfo {author} {\bibfnamefont {Matthew P.~A.}\
  \bibnamefont {Fisher}}, \bibinfo {author} {\bibfnamefont {Peter~B.}\
  \bibnamefont {Weichman}}, \bibinfo {author} {\bibfnamefont {G.}~\bibnamefont
  {Grinstein}}, \ and\ \bibinfo {author} {\bibfnamefont {Daniel~S.}\
  \bibnamefont {Fisher}},\ }\bibfield  {title} {\enquote {\bibinfo {title}
  {Boson localization and the superfluid-insulator transition},}\ }\href
  {\doibase 10.1103/PhysRevB.40.546} {\bibfield  {journal} {\bibinfo  {journal}
  {Phys. Rev. B}\ }\textbf {\bibinfo {volume} {40}},\ \bibinfo {pages}
  {546--570} (\bibinfo {year} {1989})}\BibitemShut {NoStop}%
\bibitem [{\citenamefont {Sheshadri}\ \emph {et~al.}(1993)\citenamefont
  {Sheshadri}, \citenamefont {Krishnamurthy}, \citenamefont {Pandit},\ and\
  \citenamefont {Ramakrishnan}}]{SheshadriEPL1993}%
  \BibitemOpen
  \bibfield  {author} {\bibinfo {author} {\bibfnamefont {K.}~\bibnamefont
  {Sheshadri}}, \bibinfo {author} {\bibfnamefont {H.~R.}\ \bibnamefont
  {Krishnamurthy}}, \bibinfo {author} {\bibfnamefont {R.}~\bibnamefont
  {Pandit}}, \ and\ \bibinfo {author} {\bibfnamefont {T.~V.}\ \bibnamefont
  {Ramakrishnan}},\ }\bibfield  {title} {\enquote {\bibinfo {title} {Superfluid
  and insulating phases in an interacting-boson model: Mean-field theory and
  the rpa},}\ }\href {\doibase 10.1209/0295-5075/22/4/004} {\bibfield
  {journal} {\bibinfo  {journal} {Europhysics Letters}\ }\textbf {\bibinfo
  {volume} {22}},\ \bibinfo {pages} {257} (\bibinfo {year} {1993})}\BibitemShut
  {NoStop}%
\bibitem [{\citenamefont {Freericks}\ and\ \citenamefont
  {Monien}(1994)}]{FreericksEPL1994}%
  \BibitemOpen
  \bibfield  {author} {\bibinfo {author} {\bibfnamefont {J.~K.}\ \bibnamefont
  {Freericks}}\ and\ \bibinfo {author} {\bibfnamefont {H.}~\bibnamefont
  {Monien}},\ }\bibfield  {title} {\enquote {\bibinfo {title} {Phase diagram of
  the bose-hubbard model},}\ }\href {\doibase 10.1209/0295-5075/26/7/012}
  {\bibfield  {journal} {\bibinfo  {journal} {Europhysics Letters}\ }\textbf
  {\bibinfo {volume} {26}},\ \bibinfo {pages} {545} (\bibinfo {year}
  {1994})}\BibitemShut {NoStop}%
\bibitem [{\citenamefont {Jaksch}\ \emph {et~al.}(1998)\citenamefont {Jaksch},
  \citenamefont {Bruder}, \citenamefont {Cirac}, \citenamefont {Gardiner},\
  and\ \citenamefont {Zoller}}]{JakschPRL1998}%
  \BibitemOpen
  \bibfield  {author} {\bibinfo {author} {\bibfnamefont {D.}~\bibnamefont
  {Jaksch}}, \bibinfo {author} {\bibfnamefont {C.}~\bibnamefont {Bruder}},
  \bibinfo {author} {\bibfnamefont {J.~I.}\ \bibnamefont {Cirac}}, \bibinfo
  {author} {\bibfnamefont {C.~W.}\ \bibnamefont {Gardiner}}, \ and\ \bibinfo
  {author} {\bibfnamefont {P.}~\bibnamefont {Zoller}},\ }\bibfield  {title}
  {\enquote {\bibinfo {title} {Cold bosonic atoms in optical lattices},}\
  }\href {\doibase 10.1103/PhysRevLett.81.3108} {\bibfield  {journal} {\bibinfo
   {journal} {Phys. Rev. Lett.}\ }\textbf {\bibinfo {volume} {81}},\ \bibinfo
  {pages} {3108--3111} (\bibinfo {year} {1998})}\BibitemShut {NoStop}%
\bibitem [{\citenamefont {Elstner}\ and\ \citenamefont
  {Monien}(1999)}]{ElstnerPRB1999}%
  \BibitemOpen
  \bibfield  {author} {\bibinfo {author} {\bibfnamefont {N.}~\bibnamefont
  {Elstner}}\ and\ \bibinfo {author} {\bibfnamefont {H.}~\bibnamefont
  {Monien}},\ }\bibfield  {title} {\enquote {\bibinfo {title} {Dynamics and
  thermodynamics of the bose-hubbard model},}\ }\href {\doibase
  10.1103/PhysRevB.59.12184} {\bibfield  {journal} {\bibinfo  {journal} {Phys.
  Rev. B}\ }\textbf {\bibinfo {volume} {59}},\ \bibinfo {pages} {12184--12187}
  (\bibinfo {year} {1999})}\BibitemShut {NoStop}%
\bibitem [{\citenamefont {Greiner}\ \emph {et~al.}(2002)\citenamefont
  {Greiner}, \citenamefont {Mandel}, \citenamefont {Esslinger}, \citenamefont
  {H{\"a}nsch},\ and\ \citenamefont {Bloch}}]{GreinerNature2002}%
  \BibitemOpen
  \bibfield  {author} {\bibinfo {author} {\bibfnamefont {Markus}\ \bibnamefont
  {Greiner}}, \bibinfo {author} {\bibfnamefont {Olaf}\ \bibnamefont {Mandel}},
  \bibinfo {author} {\bibfnamefont {Tilman}\ \bibnamefont {Esslinger}},
  \bibinfo {author} {\bibfnamefont {Theodor~W.}\ \bibnamefont {H{\"a}nsch}}, \
  and\ \bibinfo {author} {\bibfnamefont {Immanuel}\ \bibnamefont {Bloch}},\
  }\bibfield  {title} {\enquote {\bibinfo {title} {Quantum phase transition
  from a superfluid to a mott insulator in a gas of ultracold atoms},}\ }\href
  {\doibase 10.1038/415039a} {\bibfield  {journal} {\bibinfo  {journal}
  {Nature}\ }\textbf {\bibinfo {volume} {415}},\ \bibinfo {pages} {39--44}
  (\bibinfo {year} {2002})}\BibitemShut {NoStop}%
\bibitem [{\citenamefont {Greentree}\ \emph {et~al.}(2006)\citenamefont
  {Greentree}, \citenamefont {Tahan}, \citenamefont {Cole},\ and\ \citenamefont
  {Hollenberg}}]{GreentreeNatPhys2006}%
  \BibitemOpen
  \bibfield  {author} {\bibinfo {author} {\bibfnamefont {Andrew~D}\
  \bibnamefont {Greentree}}, \bibinfo {author} {\bibfnamefont {Charles}\
  \bibnamefont {Tahan}}, \bibinfo {author} {\bibfnamefont {Jared~H}\
  \bibnamefont {Cole}}, \ and\ \bibinfo {author} {\bibfnamefont {Lloyd~CL}\
  \bibnamefont {Hollenberg}},\ }\bibfield  {title} {\enquote {\bibinfo {title}
  {Quantum phase transitions of light},}\ }\href@noop {} {\bibfield  {journal}
  {\bibinfo  {journal} {Nature Physics}\ }\textbf {\bibinfo {volume} {2}},\
  \bibinfo {pages} {856--861} (\bibinfo {year} {2006})}\BibitemShut {NoStop}%
\bibitem [{\citenamefont {van Oosten}\ \emph {et~al.}(2001)\citenamefont {van
  Oosten}, \citenamefont {van~der Straten},\ and\ \citenamefont
  {Stoof}}]{vanOostenPRA2001}%
  \BibitemOpen
  \bibfield  {author} {\bibinfo {author} {\bibfnamefont {D.}~\bibnamefont {van
  Oosten}}, \bibinfo {author} {\bibfnamefont {P.}~\bibnamefont {van~der
  Straten}}, \ and\ \bibinfo {author} {\bibfnamefont {H.~T.~C.}\ \bibnamefont
  {Stoof}},\ }\bibfield  {title} {\enquote {\bibinfo {title} {Quantum phases in
  an optical lattice},}\ }\href {\doibase 10.1103/PhysRevA.63.053601}
  {\bibfield  {journal} {\bibinfo  {journal} {Phys. Rev. A}\ }\textbf {\bibinfo
  {volume} {63}},\ \bibinfo {pages} {053601} (\bibinfo {year}
  {2001})}\BibitemShut {NoStop}%
\bibitem [{\citenamefont {Barmettler}\ \emph {et~al.}(2012)\citenamefont
  {Barmettler}, \citenamefont {Poletti}, \citenamefont {Cheneau},\ and\
  \citenamefont {Kollath}}]{BarmettlerPRA2012}%
  \BibitemOpen
  \bibfield  {author} {\bibinfo {author} {\bibfnamefont {Peter}\ \bibnamefont
  {Barmettler}}, \bibinfo {author} {\bibfnamefont {Dario}\ \bibnamefont
  {Poletti}}, \bibinfo {author} {\bibfnamefont {Marc}\ \bibnamefont {Cheneau}},
  \ and\ \bibinfo {author} {\bibfnamefont {Corinna}\ \bibnamefont {Kollath}},\
  }\bibfield  {title} {\enquote {\bibinfo {title} {Propagation front of
  correlations in an interacting bose gas},}\ }\href {\doibase
  10.1103/PhysRevA.85.053625} {\bibfield  {journal} {\bibinfo  {journal} {Phys.
  Rev. A}\ }\textbf {\bibinfo {volume} {85}},\ \bibinfo {pages} {053625}
  (\bibinfo {year} {2012})}\BibitemShut {NoStop}%
\bibitem [{\citenamefont {Talukdar}\ and\ \citenamefont
  {Blume}(2022)}]{TalukdarPRA2022}%
  \BibitemOpen
  \bibfield  {author} {\bibinfo {author} {\bibfnamefont {J.}~\bibnamefont
  {Talukdar}}\ and\ \bibinfo {author} {\bibfnamefont {D.}~\bibnamefont
  {Blume}},\ }\bibfield  {title} {\enquote {\bibinfo {title} {Undamped rabi
  oscillations due to polaron-emitter hybrid states in a nonlinear photonic
  waveguide coupled to emitters},}\ }\href {\doibase
  10.1103/PhysRevA.106.013722} {\bibfield  {journal} {\bibinfo  {journal}
  {Phys. Rev. A}\ }\textbf {\bibinfo {volume} {106}},\ \bibinfo {pages}
  {013722} (\bibinfo {year} {2022})}\BibitemShut {NoStop}%
\bibitem [{\citenamefont {Karnieli}\ \emph {et~al.}(2024)\citenamefont
  {Karnieli}, \citenamefont {Tziperman}, \citenamefont {Roques-Carmes},\ and\
  \citenamefont {Fan}}]{KarnieliArXiv2024}%
  \BibitemOpen
  \bibfield  {author} {\bibinfo {author} {\bibfnamefont {Aviv}\ \bibnamefont
  {Karnieli}}, \bibinfo {author} {\bibfnamefont {Offek}\ \bibnamefont
  {Tziperman}}, \bibinfo {author} {\bibfnamefont {Charles}\ \bibnamefont
  {Roques-Carmes}}, \ and\ \bibinfo {author} {\bibfnamefont {Shanhui}\
  \bibnamefont {Fan}},\ }\href {https://arxiv.org/abs/2405.20241} {\enquote
  {\bibinfo {title} {Decoherence-free many-body hamiltonians in nonlinear
  waveguide quantum electrodynamics},}\ } (\bibinfo {year} {2024}),\ \Eprint
  {http://arxiv.org/abs/2405.20241} {arXiv:2405.20241 [quant-ph]} \BibitemShut
  {NoStop}%
\bibitem [{\citenamefont {Caleffi}\ \emph {et~al.}(2023)\citenamefont
  {Caleffi}, \citenamefont {Capone},\ and\ \citenamefont
  {Carusotto}}]{CaleffiPRL2023}%
  \BibitemOpen
  \bibfield  {author} {\bibinfo {author} {\bibfnamefont {Fabio}\ \bibnamefont
  {Caleffi}}, \bibinfo {author} {\bibfnamefont {Massimo}\ \bibnamefont
  {Capone}}, \ and\ \bibinfo {author} {\bibfnamefont {Iacopo}\ \bibnamefont
  {Carusotto}},\ }\bibfield  {title} {\enquote {\bibinfo {title} {Collective
  excitations of a strongly correlated nonequilibrium photon fluid across the
  insulator-superfluid phase transition},}\ }\href {\doibase
  10.1103/PhysRevLett.131.193604} {\bibfield  {journal} {\bibinfo  {journal}
  {Phys. Rev. Lett.}\ }\textbf {\bibinfo {volume} {131}},\ \bibinfo {pages}
  {193604} (\bibinfo {year} {2023})}\BibitemShut {NoStop}%
\bibitem [{\citenamefont {Pitaevskii}\ and\ \citenamefont
  {Stringari}(2016)}]{pitaevskii2016bose}%
  \BibitemOpen
  \bibfield  {author} {\bibinfo {author} {\bibfnamefont {Lev}\ \bibnamefont
  {Pitaevskii}}\ and\ \bibinfo {author} {\bibfnamefont {Sandro}\ \bibnamefont
  {Stringari}},\ }\href@noop {} {\emph {\bibinfo {title} {Bose-Einstein
  condensation and superfluidity}}},\ Vol.\ \bibinfo {volume} {164}\ (\bibinfo
  {publisher} {Oxford University Press},\ \bibinfo {year} {2016})\BibitemShut
  {NoStop}%
\bibitem [{\citenamefont {K\"uhner}\ and\ \citenamefont
  {Monien}(1998)}]{KuhnerPRB1998}%
  \BibitemOpen
  \bibfield  {author} {\bibinfo {author} {\bibfnamefont {T.~D.}\ \bibnamefont
  {K\"uhner}}\ and\ \bibinfo {author} {\bibfnamefont {H.}~\bibnamefont
  {Monien}},\ }\bibfield  {title} {\enquote {\bibinfo {title} {Phases of the
  one-dimensional bose-hubbard model},}\ }\href {\doibase
  10.1103/PhysRevB.58.R14741} {\bibfield  {journal} {\bibinfo  {journal} {Phys.
  Rev. B}\ }\textbf {\bibinfo {volume} {58}},\ \bibinfo {pages}
  {R14741--R14744} (\bibinfo {year} {1998})}\BibitemShut {NoStop}%
\bibitem [{\citenamefont {Stoof}\ \emph {et~al.}(2009)\citenamefont {Stoof},
  \citenamefont {Gubbels},\ and\ \citenamefont
  {Dickerscheid}}]{stoof2009ultracold}%
  \BibitemOpen
  \bibfield  {author} {\bibinfo {author} {\bibfnamefont {Henk~TC}\ \bibnamefont
  {Stoof}}, \bibinfo {author} {\bibfnamefont {Koos~B}\ \bibnamefont {Gubbels}},
  \ and\ \bibinfo {author} {\bibfnamefont {Dennis}\ \bibnamefont
  {Dickerscheid}},\ }\href@noop {} {\emph {\bibinfo {title} {Ultracold quantum
  fields}}}\ (\bibinfo  {publisher} {Springer},\ \bibinfo {year}
  {2009})\BibitemShut {NoStop}%
\bibitem [{Sup()}]{SupplMat}%
  \BibitemOpen
  \href@noop {} {}\bibinfo {howpublished} {See Supplemental Material at [URL]
  for detailed derivations of the Bose-Hubbard quasiparticle descriptions, the
  Markovian master equation, collective decay rates, and coherent interactions,
  which includes Ref.~\cite{breuer2002theory}.}\BibitemShut {Stop}%
\bibitem [{\citenamefont {Altman}\ and\ \citenamefont
  {Auerbach}(2002)}]{AltmanPRL2002}%
  \BibitemOpen
  \bibfield  {author} {\bibinfo {author} {\bibfnamefont {Ehud}\ \bibnamefont
  {Altman}}\ and\ \bibinfo {author} {\bibfnamefont {Assa}\ \bibnamefont
  {Auerbach}},\ }\bibfield  {title} {\enquote {\bibinfo {title} {Oscillating
  superfluidity of bosons in optical lattices},}\ }\href {\doibase
  10.1103/PhysRevLett.89.250404} {\bibfield  {journal} {\bibinfo  {journal}
  {Phys. Rev. Lett.}\ }\textbf {\bibinfo {volume} {89}},\ \bibinfo {pages}
  {250404} (\bibinfo {year} {2002})}\BibitemShut {NoStop}%
\bibitem [{\citenamefont {Huber}\ \emph {et~al.}(2007)\citenamefont {Huber},
  \citenamefont {Altman}, \citenamefont {B\"uchler},\ and\ \citenamefont
  {Blatter}}]{HuberPRB2007}%
  \BibitemOpen
  \bibfield  {author} {\bibinfo {author} {\bibfnamefont {S.~D.}\ \bibnamefont
  {Huber}}, \bibinfo {author} {\bibfnamefont {E.}~\bibnamefont {Altman}},
  \bibinfo {author} {\bibfnamefont {H.~P.}\ \bibnamefont {B\"uchler}}, \ and\
  \bibinfo {author} {\bibfnamefont {G.}~\bibnamefont {Blatter}},\ }\bibfield
  {title} {\enquote {\bibinfo {title} {Dynamical properties of ultracold bosons
  in an optical lattice},}\ }\href {\doibase 10.1103/PhysRevB.75.085106}
  {\bibfield  {journal} {\bibinfo  {journal} {Phys. Rev. B}\ }\textbf {\bibinfo
  {volume} {75}},\ \bibinfo {pages} {085106} (\bibinfo {year}
  {2007})}\BibitemShut {NoStop}%
\bibitem [{\citenamefont {Cheneau}\ \emph {et~al.}(2012)\citenamefont
  {Cheneau}, \citenamefont {Barmettler}, \citenamefont {Poletti}, \citenamefont
  {Endres}, \citenamefont {Schau{\ss}}, \citenamefont {Fukuhara}, \citenamefont
  {Gross}, \citenamefont {Bloch}, \citenamefont {Kollath},\ and\ \citenamefont
  {Kuhr}}]{CheneauNature2012}%
  \BibitemOpen
  \bibfield  {author} {\bibinfo {author} {\bibfnamefont {Marc}\ \bibnamefont
  {Cheneau}}, \bibinfo {author} {\bibfnamefont {Peter}\ \bibnamefont
  {Barmettler}}, \bibinfo {author} {\bibfnamefont {Dario}\ \bibnamefont
  {Poletti}}, \bibinfo {author} {\bibfnamefont {Manuel}\ \bibnamefont
  {Endres}}, \bibinfo {author} {\bibfnamefont {Peter}\ \bibnamefont
  {Schau{\ss}}}, \bibinfo {author} {\bibfnamefont {Takeshi}\ \bibnamefont
  {Fukuhara}}, \bibinfo {author} {\bibfnamefont {Christian}\ \bibnamefont
  {Gross}}, \bibinfo {author} {\bibfnamefont {Immanuel}\ \bibnamefont {Bloch}},
  \bibinfo {author} {\bibfnamefont {Corinna}\ \bibnamefont {Kollath}}, \ and\
  \bibinfo {author} {\bibfnamefont {Stefan}\ \bibnamefont {Kuhr}},\ }\bibfield
  {title} {\enquote {\bibinfo {title} {Light-cone-like spreading of
  correlations in a quantum many-body system},}\ }\href {\doibase
  10.1038/nature10748} {\bibfield  {journal} {\bibinfo  {journal} {Nature}\
  }\textbf {\bibinfo {volume} {481}},\ \bibinfo {pages} {484--487} (\bibinfo
  {year} {2012})}\BibitemShut {NoStop}%
\bibitem [{\citenamefont {Jordan}\ and\ \citenamefont
  {Wigner}(1928)}]{Jordan1928}%
  \BibitemOpen
  \bibfield  {author} {\bibinfo {author} {\bibfnamefont {P.}~\bibnamefont
  {Jordan}}\ and\ \bibinfo {author} {\bibfnamefont {E.}~\bibnamefont
  {Wigner}},\ }\bibfield  {title} {\enquote {\bibinfo {title} {{\"U}ber das
  paulische {\"a}quivalenzverbot},}\ }\href {\doibase 10.1007/BF01331938}
  {\bibfield  {journal} {\bibinfo  {journal} {Zeitschrift f{\"u}r Physik}\
  }\textbf {\bibinfo {volume} {47}},\ \bibinfo {pages} {631--651} (\bibinfo
  {year} {1928})}\BibitemShut {NoStop}%
\bibitem [{\citenamefont {Batista}\ and\ \citenamefont
  {Ortiz}(2001)}]{BatistaPRL2001}%
  \BibitemOpen
  \bibfield  {author} {\bibinfo {author} {\bibfnamefont {C.~D.}\ \bibnamefont
  {Batista}}\ and\ \bibinfo {author} {\bibfnamefont {G.}~\bibnamefont
  {Ortiz}},\ }\bibfield  {title} {\enquote {\bibinfo {title} {Generalized
  jordan-wigner transformations},}\ }\href {\doibase
  10.1103/PhysRevLett.86.1082} {\bibfield  {journal} {\bibinfo  {journal}
  {Phys. Rev. Lett.}\ }\textbf {\bibinfo {volume} {86}},\ \bibinfo {pages}
  {1082--1085} (\bibinfo {year} {2001})}\BibitemShut {NoStop}%
\bibitem [{\citenamefont {Cosco}\ \emph {et~al.}(2018)\citenamefont {Cosco},
  \citenamefont {Borrelli}, \citenamefont {Mendoza-Arenas}, \citenamefont
  {Plastina}, \citenamefont {Jaksch},\ and\ \citenamefont
  {Maniscalco}}]{CoscoPRA2018}%
  \BibitemOpen
  \bibfield  {author} {\bibinfo {author} {\bibfnamefont {Francesco}\
  \bibnamefont {Cosco}}, \bibinfo {author} {\bibfnamefont {Massimo}\
  \bibnamefont {Borrelli}}, \bibinfo {author} {\bibfnamefont {Juan~Jos\'e}\
  \bibnamefont {Mendoza-Arenas}}, \bibinfo {author} {\bibfnamefont {Francesco}\
  \bibnamefont {Plastina}}, \bibinfo {author} {\bibfnamefont {Dieter}\
  \bibnamefont {Jaksch}}, \ and\ \bibinfo {author} {\bibfnamefont {Sabrina}\
  \bibnamefont {Maniscalco}},\ }\bibfield  {title} {\enquote {\bibinfo {title}
  {Bose-hubbard lattice as a controllable environment for open quantum
  systems},}\ }\href {\doibase 10.1103/PhysRevA.97.040101} {\bibfield
  {journal} {\bibinfo  {journal} {Phys. Rev. A}\ }\textbf {\bibinfo {volume}
  {97}},\ \bibinfo {pages} {040101} (\bibinfo {year} {2018})}\BibitemShut
  {NoStop}%
\bibitem [{\citenamefont {Caleffi}\ \emph {et~al.}(2021)\citenamefont
  {Caleffi}, \citenamefont {Capone}, \citenamefont {De~Vega},\ and\
  \citenamefont {Recati}}]{CaleffiNJP2021}%
  \BibitemOpen
  \bibfield  {author} {\bibinfo {author} {\bibfnamefont {Fabio}\ \bibnamefont
  {Caleffi}}, \bibinfo {author} {\bibfnamefont {Massimo}\ \bibnamefont
  {Capone}}, \bibinfo {author} {\bibfnamefont {In{\'e}s}\ \bibnamefont
  {De~Vega}}, \ and\ \bibinfo {author} {\bibfnamefont {Alessio}\ \bibnamefont
  {Recati}},\ }\bibfield  {title} {\enquote {\bibinfo {title} {Impurity
  dephasing in a bose--hubbard model},}\ }\href@noop {} {\bibfield  {journal}
  {\bibinfo  {journal} {New Journal of Physics}\ }\textbf {\bibinfo {volume}
  {23}},\ \bibinfo {pages} {033018} (\bibinfo {year} {2021})}\BibitemShut
  {NoStop}%
\bibitem [{\citenamefont {Masson}\ and\ \citenamefont
  {Asenjo-Garcia}(2022)}]{MassonNatCommun2022}%
  \BibitemOpen
  \bibfield  {author} {\bibinfo {author} {\bibfnamefont {Stuart~J.}\
  \bibnamefont {Masson}}\ and\ \bibinfo {author} {\bibfnamefont {Ana}\
  \bibnamefont {Asenjo-Garcia}},\ }\bibfield  {title} {\enquote {\bibinfo
  {title} {Universality of dicke superradiance in arrays of quantum
  emitters},}\ }\href {\doibase 10.1038/s41467-022-29805-4} {\bibfield
  {journal} {\bibinfo  {journal} {Nature Communications}\ }\textbf {\bibinfo
  {volume} {13}},\ \bibinfo {pages} {2285} (\bibinfo {year}
  {2022})}\BibitemShut {NoStop}%
\bibitem [{\citenamefont {Recati}\ \emph {et~al.}(2005)\citenamefont {Recati},
  \citenamefont {Fedichev}, \citenamefont {Zwerger}, \citenamefont {von
  Delft},\ and\ \citenamefont {Zoller}}]{RecatiPRL2005}%
  \BibitemOpen
  \bibfield  {author} {\bibinfo {author} {\bibfnamefont {A.}~\bibnamefont
  {Recati}}, \bibinfo {author} {\bibfnamefont {P.~O.}\ \bibnamefont
  {Fedichev}}, \bibinfo {author} {\bibfnamefont {W.}~\bibnamefont {Zwerger}},
  \bibinfo {author} {\bibfnamefont {J.}~\bibnamefont {von Delft}}, \ and\
  \bibinfo {author} {\bibfnamefont {P.}~\bibnamefont {Zoller}},\ }\bibfield
  {title} {\enquote {\bibinfo {title} {Atomic quantum dots coupled to a
  reservoir of a superfluid bose-einstein condensate},}\ }\href {\doibase
  10.1103/PhysRevLett.94.040404} {\bibfield  {journal} {\bibinfo  {journal}
  {Phys. Rev. Lett.}\ }\textbf {\bibinfo {volume} {94}},\ \bibinfo {pages}
  {040404} (\bibinfo {year} {2005})}\BibitemShut {NoStop}%
\bibitem [{\citenamefont {Breuer}\ and\ \citenamefont
  {Petruccione}(2002)}]{breuer2002theory}%
  \BibitemOpen
  \bibfield  {author} {\bibinfo {author} {\bibfnamefont {Heinz-Peter}\
  \bibnamefont {Breuer}}\ and\ \bibinfo {author} {\bibfnamefont {Francesco}\
  \bibnamefont {Petruccione}},\ }\href@noop {} {\emph {\bibinfo {title} {The
  theory of open quantum systems}}}\ (\bibinfo  {publisher} {OUP Oxford},\
  \bibinfo {year} {2002})\BibitemShut {NoStop}%
\end{thebibliography}%

\end{document}